\documentclass[epj]{svjour}
\usepackage{latexsym}
\usepackage{graphics,epsfig,subfigure}

\begin{document}

\title{Renormalization by Continuous Unitary Transformations:
One-Dimensional Spinless Fermions}
%\subtitle{}
\author{Caspar P. Heidbrink \and G\"otz S. Uhrig
%\thanks{\emph{Present address:}}%
}

\authorrunning{C.P.~Heidbrink,  G.S.~Uhrig}
\titlerunning{Renormalization by Continuous Unitary Transformations}

\institute{ Institut f\"ur Theoretische Physik, 
Universit\"at zu K\"oln, Z\"ulpicher Stra{\ss}e 77, 50937 K\"oln}
\date{Received: date / Revised version: date}

\abstract{A renormalization scheme for interacting fermionic systems
is presented where the renormalization is carried out in terms of the
fermionic degrees of freedom. 
The scheme is based on continuous unitary transformations of
the hamiltonian which stays hermitian throughout the renormalization flow,
whereby any frequency dependence is avoided. The approach is illustrated
in detail for a model of spinless fermions with nearest neighbour
repulsion in one dimension. Even though the fermionic degrees of freedom
do not provide an easy starting point in one dimension favorable results
are obtained which agree well with the exact findings based on Bethe ansatz.
\PACS{
  {05.10.Cc}{Renormalization group methods}   \and
  {71.10.Pm}{Fermions in reduced dimensions}   \and
  {71.10.Ay}{Fermi-liquid theory and other phenomenological models}
  }
}

\maketitle

\section{Introduction}
\label{intro}

With the discovery of high temperature superconductivity
in layered cuprates of perovskite type
\cite{bedno86} and the subsequent most intensive theoretical considerations 
 of this intriguing phenomenon (see e.g.\ Refs.~\cite{ander97,carls02})
it has become
apparent that a reliable approach to strongly interacting
fermionic systems is lacking. 
So the efforts to formulate the very successful
concept of renormalization \cite{wilso75} also for extended
interacting fermionic systems have been intensified considerably
during the last decade, see e.g.\ Ref.~\cite{shank94}. 
By now the literature on this
approach is so wide that it is not possible to provide an exhaustive
list. This  underlines the importance that is accorded to 
this topic.

So far a number of renormalizing schemes has been applied to
two-dimensional Hubbard-type models which are relevant for
high temperature superconductivity \cite{zanch97,halbo00a,honer01a}.
These schemes rely conceptually on diagrammatic perturbation theory in the
interaction strength. 
%%% NEU
They are non-perturbative in the sense that
infinite orders of the interaction are kept; the necessary 
truncation concerns 
terms of a certain structure, e.g.\ six points correlations, not
terms of a certain order in the interaction.
%%% NEU
The Fermi surface is discretized and the various
scattering couplings across the Fermi sea are suitably parametrized.
It is possible to detect whether or not
the renormalized couplings decrease of diverge in the course of the flow.
In this way, robust evidence for the occurrence of $d$-wave superconductivity
was found. The corresponding couplings diverge for certain values of doping
and interaction. 
%%% NEU
So far, however, no calculations exist for
dynamical correlations like the ARPES response
at finite energies and wave vectors (measured
relative to the Fermi surface) as required for the
understanding of the experimental findings. This is due to the 
great complexity of the problem which requires --
among other difficulties --  to follow the flow of the
observables as well, see for instance the appendix in Ref.\ \cite{honer01a}.
For an impurity in a spinless Luttinger liquid a spectral density
at all energies 
was computed by a one-particle irreducible renormalization approach
neglecting, however, the renormalization of the two-particle vertex
\cite{meden02}.
%%% NEU

Starting from a flow equation approach as proposed by Wegner
\cite{wegne94} and quite similarly by G{\l}azeck  and Wilson 
\cite{glaze93,glaze94} a renormalizing scheme based on continuous
unitary transformations (CUTs) has been proposed \cite{heidb02a}.
By an appropriately chosen unitary transformation an effective hamiltonian
$H_{\rm eff}$
is obtained. The transformation is tuned in such a way that $H_{\rm eff}$
conserves the number of quasi-particles. Due to this property 
the calculation of dynamical correlation function for all energies
becomes possible
as was shown for spin ladders \cite{knett01b}. For this 
reason, we consider the renormalizing CUT approach to have a
particularly great potential. This expectation is supported
decisively by  recent work on quantum chemical systems \cite{white02} 
where White could show that the numerical 
application of a continuous unitary transformation similar
to the one used in Ref.~\cite{heidb02a} and here leads to excellent results.

It is the aim of the present work to explain the technical details of
the calculations announced in Ref.~\cite{heidb02a}. To this end,
intermediate results will be shown. We hope that the available data
will make it possible to conceive also analytical treatments  which
capture the  essential physics. The medium-term objective is to 
generalize the renormalizing CUT approach to more realistic models.
A demanding challenge is to treat the dimensional crossover between
one- and two-dimensional interacting fermionic models. This is
of great experimental relevance since the physical systems existing
in nature are at best quasi-one-dimensional, i.e.\
strongly anisotropic, so that the higher dimensionality enters always
at a certain stage.

Two dimensions are the most demanding case for the theoretical description
of strong correlations. In three dimensions the powerful Fermi liquid
theory is well established which is based on the observation that
the quasi-particles as such yield a good description of the low-lying
excitations (see e.g.\ \cite{nozie97}). 
The interaction {\em between} the quasi-particles is 
not essential. In one dimension on the other hand, the 
collective plasmon modes dominate the low-energy physics completely
so that the fermionic hamiltonian can be mapped to a bosonic one
representing so-called Luttinger liquids
(see e.g.\ \cite{voit95,gogol98}). This phenomenon can be seen
as a binding (or anti-binding) of a pair consisting of a hole and
a fermion. This implies that it is a signature of a dominating interaction 
between the quasi-particles. This fact is commonly interpreted
as the failure of a description of the energetically low-lying 
physics in terms of quasi-particles. 

%%% NEU
Considering two dimensional strongly interacting systems it is
shown that they are generically Fermi liquids in the weak
coupling regime \cite{feldm95}. So, qualitatively, two dimensional
systems are similar to three dimensional ones in the weak coupling limit.
 But at strong coupling this is no longer true. For given generic hopping
element $t$ and interaction strength $U$ the lower dimensional system
is more  influenced by strong correlations than the corresponding higher
dimensional one. This stems from two effects: (i) The band width being 
proportional to the coordination number
is higher in the higher dimensional case. (ii) If collective modes,
damped or undamped, are formed
their density of states (DOS) at low energies is higher in lower dimensions.
For instance, the DOS $\rho(\omega)$ of linear dispersing modes 
$\omega \propto |{\bf k}|$ behaves like $\rho(\omega)\propto \omega^{d-1}$
with dimension $d$.

In the above sense,  two dimensional strongly interacting systems
represent an intermediate situation. Generically, neither the interactions
between the quasi-particles  can be sufficiently described by a
Landau function as in three dimensions,
 nor is the physics {\em completely} dominated by collective modes
formed from bound particle-hole pairs. Hence, in two dimensions
one has to have a theoretical tool which is able to reconcile both
main features: 
collective modes occur and they are important, but they do not
exhaust all degrees of freedom. There are also quasi-particles. But
their interaction is very important. 

%%% NEU
The above considerations are the motivation to  show here that
it is possible to use continuous unitary transformations and quasi-particle
description
%%% NEU
 for one dimensional systems to recover the known results. In particular,
we will look at the momentum distribution in the ground state which differs
significantly between Fermi liquids and Luttinger liquids. In 
Fermi liquids a jump by $Z_{k_{\rm F}}$, the quasi-particle weight,
occurs at the Fermi level whereas in Luttinger liquids only a power-law
behaviour occurs \cite{voit95,gogol98}. 
%%% NEU
Our investigation is intended 
to be a test case for the method, not as a means to obtain new and so far
unknown data. 
%%% NEU
Evidence is provided that in one dimension
the physics has not turned bosonic out of the blue but that a description
in terms of fermionic 
quasi-particles is still reasonable even though a description
in terms of bosons is easier. In view of the medium-term aim
to describe the dimensional crossover we consider
the fermionic approach to be  a necessary prerequisite.

The paper is set-up as follows. After this Introduction the method
 is described in Sect.~2. Also the model to which the 
continuous unitary transformation is applied is given in detail. In
Sect.~3 the numerical results are shown and discussed. The 
comprehensive Discussion concludes the article in Sect.~4.

\section{Method and Model}
\label{sec:1}
In general, we consider a translationally invariant 
system of $N$ interacting fermions
\begin{eqnarray}
H&=&N  E+\sum_{k} 
\varepsilon_k :c^{\dagger}_{k}c^{\phantom\dagger}_{k}:
\nonumber\\&&
+\frac{1}{N}\sum_{kqp}
\Gamma_{kqp}:c^{\dagger}_{k+q}c^{\dagger}_{k-q}
c^{\phantom\dagger}_{k-p}
c^{\phantom\dagger}_{k+p}:\ .
\label{general_ham}
\end{eqnarray}
Here $c^\dagger_k$ ($c^{\phantom\dagger}_{k}$) creates (annihilates) a
fermion at wave vector $k$ in momentum space. Note that the parametrization
of the scattering processes is not the conventional one. Our choice, however,
is more apt to represent the inherent symmetries (see below).
The fermions appearing are considered to be spinless, i.e.\ there  is at
maximum one per site and no spin index occurs.
The colons $:\ldots :$ denote normal ordering with respect to the 
non-interacting Fermi sea.  This commonly
used classification makes it easier to trace the effect of
the individual terms (cf.~appendix A). 
The function $\Gamma_{kqp}$ is the vertex function embodying the
amplitudes for all possible scattering processes.

\subsection{Method}
\label{method}
The method which we use to analyse the hamiltonian (\ref{general_ham}),
 is the method of continuous unitary transformations (CUTs)  based on
the flow equation approach  proposed by Wegner \cite{wegne94}. 
The unitary transformation $U(\ell)$ is parametrized by the continuous
 variable $\ell$ ranging between $0$ and $\infty$. A
 given hamiltonian $H(0)$ is continuously mapped onto an effective model 
$H_{\rm eff}$ as $\ell$ is taken from zero to infinity.
The continuous unitary 
transformation is defined locally in $\ell$ by the antihermitian generator 
$\eta(\ell)$
\begin{equation}
\frac{d}{dl}H(\ell)=\left[\eta(\ell),H(\ell)\right]\ .
\label{general_fe}
\end{equation}
Of course, all other observables $O$ have to be subject to 
the same transformations
\begin{equation}
\frac{d}{dl}O(\ell)=\left[\eta(\ell),O(\ell)\right]
\label{operator_fe}
\end{equation}
since the expectation values and correlations shall not be altered
by the transformation.

The main task is to determine $\eta(\ell)=\eta(H(\ell))$ in a way that
 brings the hamiltonian  systematically closer to a simpler structure. 
Once the
 generator is chosen 
and the commutator calculated, one obtains a high dimensional
 set of coupled ordinary differential equations. These have to be
solved in the limit $\ell\to \infty$. 
\begin{figure}
\centering
\epsfig{width=3cm,figure=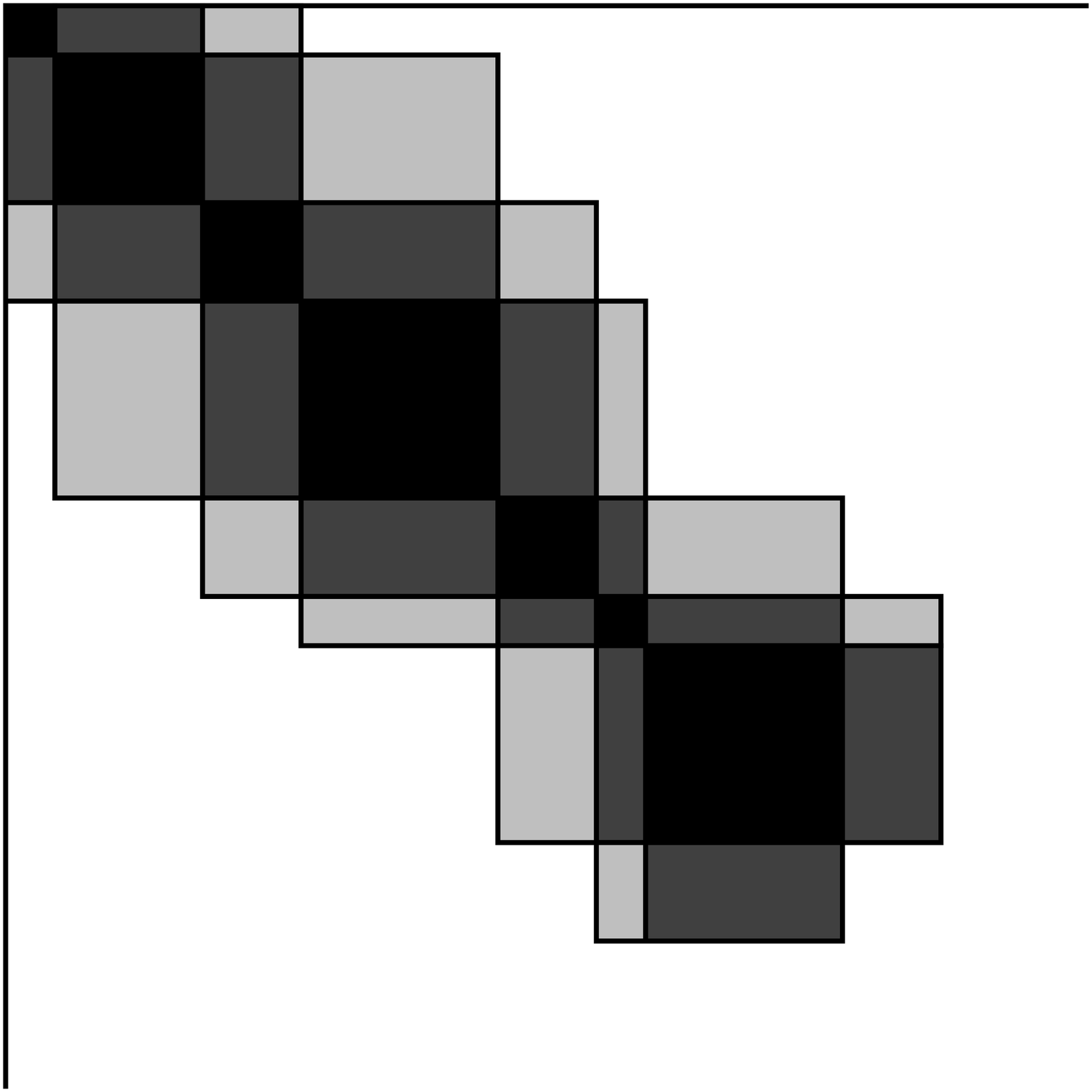}
\raisebox{1.3cm}{
$\Longrightarrow$
}
\epsfig{width=3cm,figure=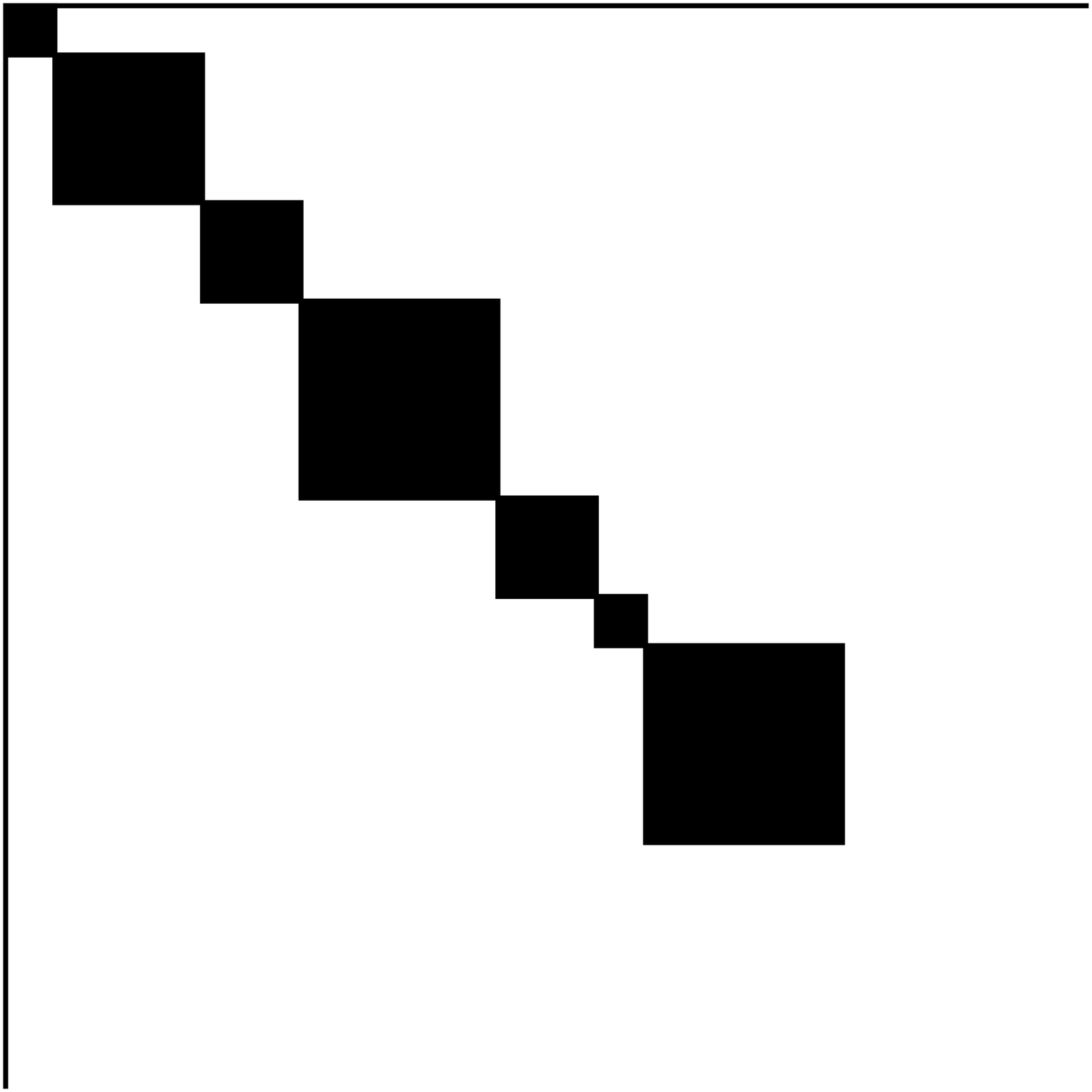}
\caption{Hamiltonian with a block band structure transformed into an 
effective block diagonal hamiltonian. Each black block stands for the 
action of the hamiltonian in a subspace of the Hilbert space with a 
given number of excitations. Let us assume that the uppermost block
stands for the vacuum (no excitation); the next lower block acts
on states with one single excitation; the next lower block acts on 
states with two excitations and so on. In the example depicted the
number of excitations can change at most by two
on a single application of the hamiltonian.
 This is the block band structure.}
\label{block_band}
\end{figure}

For the choice of $\eta(\ell)$ we focus on the case where
 the original hamiltonian $H$ has a so called 
\emph{block band structure} with respect to a certain counting
operator $Q$, i.e.\ an operator with a spectrum of non-negative
integers. The operator $Q$ shall be the operator
counting the number of excitations present in the system. The ground
state $|0\rangle$ of $Q$, the so-called 
quasi-particle vacuum,   is the state without any excitation
so that
\begin{equation}
Q|0\rangle = 0
\end{equation}
holds. The block band 
structure occurs if the whole hamiltonian $H$ changes the counting operator
$Q$ at maximum by a certain value $M < \infty$. This means that
$H$ links two eigen states $|i\rangle$ and $|j\rangle$ of $Q$ only if their 
eigen values $q_i$ and $q_j$ differ at most by $M$: $|q_i - q_j|\le M$.
In Fig.~\ref{block_band} the case $M=2$ is illustrated.

In view of systems of interacting fermions (\ref{general_ham})
we choose $Q$ to be the operator counting
the number of quasi-particles, i.e.\ the number of 
particles above the Fermi level plus the number of holes below
\begin{equation}
\label{count-operator}
Q = \sum_k \mbox{sign}(|k|-k_{\rm F}) 
:c^{\dagger}_{k}c^{\phantom\dagger}_{k}: \ .
\end{equation}
This choice implies that we intend to use ordinary quasi-particles
as the elementary excitations. Generally, we like to stress that the
choice of $Q$ requires some physical intuition or a certain presumption
about the problem under study. The choice of $Q$ can be considered as the
choice of a starting point. Depending on the quality of this choice 
subsequent approximations will be more or less reliable. This point will
be discussed below in more detail.

First, however, we turn to the choice of the generator.
 For band-diagonal matrices Mielke proposed an ansatz \cite{mielk98} 
that was generalized to many-body problems with block-band structures by 
Knetter and Uhrig \cite{knett00a}. 
The antihermitian generator $\eta(\ell)$  is chosen such that
its matrix elements are given in an eigen basis of $Q$ by
\begin{equation}
  \eta_{ij}(\ell)=\mbox{sign}(q_{i}(\ell)-q_{j}(\ell))H_{ij}(\ell)\ .
\label{mku_cuts}
\end{equation}
 For the interacting fermions as in
(\ref{general_ham}) the conventional occupation number basis is appropriate.

Since in general the eigen space for a given number of 
excitations $q_i$ has a finite dimension we use the convention that $A_{ij}$
is not only a single matrix element but the whole submatrix of $A$ which
 connects the eigen space belonging to $q_j$ (domain) to the eigen space 
belonging to $q_i$ (co-domain). With this convention Eq.~\ref{mku_cuts}
becomes a matrix equation. Inserting Eq.~\ref{mku_cuts} into the general
flow equation (\ref{general_fe}) yields
\begin{eqnarray}
\nonumber
&&\frac{d}{dl}H_{ij}=
-\mbox{sign}(q_{i}-q_{j})(H_{ii}H_{ij}-H_{ij}H_{jj})\\
&&\quad
+\sum_{k\neq ij}\left(\mbox{sign}(q_{i}-q_{k})+\mbox{sign}(q_{j}-q_{k})
\right)H_{ik}H_{kj}\label{flow-eq}
\end{eqnarray}
which is also a matrix equation, i.e.\ the sequence of the matrices in
the products matters. One can show \cite{mielk98,knett00a} that (\ref{flow-eq})
\begin{itemize}
\item preserves the block band structure and
\item leads to a block diagonal $H_{\rm eff}$ where
 $q_{i}\neq q_{j} \Rightarrow H_{{\rm eff},ij}=0$. The condition necessary
for these conclusions is that the spectrum of the hamiltonian
 is bounded from below which constitutes a natural assumption 
for physical systems.
\end{itemize}
The block diagonality of the effective model is equivalent to the
commutation of the effective model  with $Q$ $[Q,H_{{\rm eff}}]=0$.

Furthermore, we show that the ground state of the system is mapped onto the 
state with no elementary excitations, i.e.\ the vacuum $|0\rangle$
without any excited quasi-particles, in the course of the transformation.
Let us assume that the ground state is not degenerate. 
Without loss of generality the indexing is chosen such that 
$i=0$ refers to the vacuum with $q_0=0$. For $j=0$ we  obtain from
(\ref{flow-eq})
\begin{eqnarray}
\nonumber
\frac{d}{dl}H_{i0}&=&
-(H_{ii}H_{i0}-H_{i0}E)\\
&&+\sum_{k\neq i0}\left(\mbox{sign}(q_{i}-q_{k})-1)
\right)H_{ik}H_{k0}\ ,
\label{ground_flow}
\end{eqnarray}
where $E:=\langle 0|H|0\rangle=H_{00}$ is the energy of the
 vacuum. $H_{i0}$ is  a vector that connects the vacuum with  other 
states. For large $\ell$ the flow has led the system already close
to block diagonality \cite{mielk98,knett00a}. Then the matrices $H_{ik}$ 
linking blocks of different number of elementary excitations are small 
quantities so that Eq.~(\ref{ground_flow}) is dominated by the first
term on the right hand side. The products $H_{ik}H_{k0}$ in the sum are
smaller by one order in the off-diagonal matrices. So the asymptotic behaviour
is given by
\begin{equation}
\label{e-conver}
\frac{d}{dl}H_{i0}\approx(E-H_{ii})H_{i0}\ .
\end{equation}
Since $H_{i0}$ has to tend to zero due to the general convergence 
\cite{mielk98,knett00a} we deduce from (\ref{e-conver}) that $E-H_{ii}\leq 0$ 
for large $\ell$. This implies that $E$ is lower in energy than
 any state different from the vacuum which is linked to the vacuum by a 
non-zero element $H_{i0}$. We interprete the physical content of this result 
in the following way. In a block band diagonal hamiltonian as illustrated
on the left side in Fig.~\ref{block_band} the vacuum is linked to states
with a certain finite number (bounded by $M$) 
of elementary excitations. From (\ref{e-conver})
we learn that the vacuum is lower in energy than those states. 
Generically, this implies that also all other states which contain an 
unrestricted number of elementary excitations lie higher in energy.
We conclude that the vacuum is indeed the ground state {\em unless}
 a phase transition takes place. 

To illustrate the latter statement in more detail let us think
about a hamiltonian which is controlled by an interaction  parameter $U$ such
that the vacuum is the ground state for $U=0$ without any transformation.
Then the situation changes gradually and smoothly as $U$ is turned on
so that the ground state is mapped onto the vacuum. This is
true till a singularity occurs, that is a phase transition.
A second order phase transition would
be signaled by the softening -- vanishing of the eigen energy --
 of one of the excited states
 (for an example see the single triplet excitation in Ref.~\cite{knett00b}).
Technically, this implies that the convergence of $H_{i0}$ as given
by Eq.~(\ref{e-conver}) breaks down since the expression in the
bracket on the right hand side vanishes. This reflects the physical fact that
our approach will not work  beyond the phase transition since there the
original vacuum is no longer a good reference state for the true ground
state. But it is possible to use the approach until the phase transition
is reached and the second order phase transition can be detected by the
vanishing of the energy of one of the excited states.

First order transitions spoil also the applicability of a smooth mapping.
But they cannot be detected locally.  A first order jump occurs when a
completely different state comes down in energy. Such a completely different
state is built from a macroscopic number of elementary excitations and will
in general not be connected to the vacuum by a finite element $H_{i0}$.
This means that the smooth mapping constructed by the CUT may still work
even though a first order transition has occurred simply because there is no
local connection of the local energy minimum, which is mapped to the
vacuum, to the global one. This concludes the general considerations
about the mapping generated by the CUT between the ground state and the vacuum
of elementary excitations.

Now we  return to the model of interacting fermions in (\ref{general_ham}).
Inspection shows that this model has a block band structure with respect to
the counting operator (\ref{count-operator}) of quasi-particles. 
If a particle from below the Fermi level is scattered to a state above 
the Fermi level two elementary excitations are created: one hole and one
particle. The inverse process corresponds to the decrement of the number
of quasi-particles by two.
If the particle is scattered from below the Fermi level to below
the Fermi level  the number of quasi-particles remains constant. The
same is true if the particle is taken from above the Fermi level to another
state above this level. If we consider at maximum scattering terms built
from four fermionic operators, as it is done in Eq.~(\ref{general_ham}), 
then two fermions are scattered so that the possible changes in the number
of quasi-particles are $0, \pm2, \pm4$. So there is a natural upper bound 
$M=4$ and the system displays indeed a block band structure. 
Note that all the statements so far were independent of the dimension of
the model.

\subsection{Model}
The explicit model we consider here is a system of  one-dimensional spinless 
fermions. Haldane used it to explain the concept of Luttinger liquids 
\cite{halda81b}. Physical realizations one may think of are either 
completely polarized electrons or anisotropic spin chains which can be
rigorously mapped onto spinless fermions by means of the 
 Jordan-Wigner transformation \cite{fradk91}. A completely
different application
is the description of vicinal surfaces where the hard-core repulsion 
between different steps
is taken into account by passing to fermions \cite{joos91}.

The model is a tight-binding model where the particles are distributed
 over a one-dimensional chain of $N$ sites with antiperiodic
 boundary conditions at half filling, i.e.\ on average each site is occupied
by half a fermion. The fermions can hop to nearest neighbour sites and there 
is a  repulsive interaction $V$ between two adjacent fermions
\begin{equation}
H=\sum^N_{i=1}\left[{\textstyle\frac{1}{2}}(a^{\dagger}_{i+1}
a^{\phantom\dagger}_{i}+\mbox{h.c.})+
V \left(n_{i+1}-{\textstyle\frac{1}{2}}\right)\left(n_{i}-
{\textstyle\frac{1}{2}}\right)\right] \, ,\label{model_space}
\end{equation}
where $a^\dagger_i$ ($a^{\phantom\dagger}_{i}$) creates (annihilates) a
fermion on site $i$ in real space. 
This model is exactly solvable in the form of an anisotropic spin model. 
The  solution is due to Bethe \cite{bethe31} and 
to  Yang and Yang \cite{yang66a,yang66b}.
More results were found later 
\cite{johns73,luthe75,halda80,halda81a,halda81b}. 
In the present context it is important to know that the system is metallic for
not too large couplings, namely for 
$V\leq 1$. In this region it is the simplest realistic tight-binding model
displaying Luttinger liquid behaviour which is why we use it as
our test case. At the  value $V=1$ a continuous phase transition takes place
into a charge density wave where the fermion density is staggered. Thus
the translation symmetry of the system is spontaneously broken. This phase
displays also a charge gap and represents hence an insulator.
In the present work, however, our interest is focused on the metallic phase.

\paragraph{Representation}

\noindent
Let us first clarify the notation that we use in the following.
Transforming the real-space representation (\ref{model_space}) into 
momentum-space and normal-ordering of the hamiltonian with respect to
the non-interacting Fermi sea yields a hamiltonian of the form in 
Eq.~(\ref{general_ham}) with the coefficients
\begin{eqnarray}
E&=&-\left({\textstyle\frac{1}{\pi}+\frac{V}{2\pi}}\right)\label{me}\\
\varepsilon_k&=&-\left(1+{\textstyle\frac{2V}{\pi}}\right)\cos k\label{md}\\
\Gamma_{kqp}&=&{\phantom -}V\sin q \sin p\ .\label{mg}
\end{eqnarray}
in the thermodynamic limit $N\to\infty$.
 For finite system sizes $N<\infty$ as used below the dispersion reads
\begin{equation}
\label{finitesize-disper}
\varepsilon_k = -\left(1+{\frac{2V}{N}}\sum_{|k'|<k_{\rm F}}\cos k'\right)
\cos k\ .
\end{equation}
The terms $\frac{1}{\pi}$ and $\frac{V}{2\pi}$ in $E$ correspond to the
ground state energy of the non-interacting system and to the 
 Fock term, respectively. We are not interested in these leading order
effects but will concentrate on the correlation effects as they appear
in the subsequent orders. 

Based on the parametrization chosen in Eq.~(\ref{general_ham})
 the symmetries and notation properties
 of the system manifest themselves in a  very concise way in
the vertex function $\Gamma_{kqp}$
\begin{enumerate}
\item hermitecity leads to 
  \begin{equation}
    \Gamma_{kqp}=\Gamma_{kpq}\ ,
    \label{s1}
  \end{equation}
\item inversion symmetry leads to
  \begin{equation}
    \Gamma_{kqp}=\Gamma_{-k-q-p}
    \label{s2}
  \end{equation}
\item particle-hole symmetry leads to
  \begin{equation}
    \Gamma_{kqp}=\Gamma_{k+\pi qp}=\Gamma_{kq+\pi p+\pi}\ .
    \label{s3}
  \end{equation}
\end{enumerate}
A swap of two neighboured fermionic operators in a normal-ordered product 
 leads only to an additional minus sign. Thus there is a 
redundancy in the notation $:c_1c_2\ldots:=-:c_2c_1\ldots:$
which implies that $\Gamma_{kqp}$ is not uniquely defined by
Eq.~(\ref{general_ham}). This caveat can be remedied by  the additional
requirement of the
\begin{enumerate}
\newcounter{local}
\renewcommand{\labelenumi}{\setcounter{local}{\value{enumi}4.}}
\item notation symmetry 
  \begin{equation}
    \Gamma_{kqp}=-\Gamma_{k-qp}=-\Gamma_{kq-p}\ .
    \label{r1}
  \end{equation}
\end{enumerate}
The notation symmetry implies that the momentum dependence of the
vertex function is the one given in (\ref{mg}). Note that in our
notation the vertex function $\Gamma_{kqp}$ vanishes automatically
if fermions are scattered from the Fermi points to the Fermi points
($k=\pm\pi/2$ and $q,p \in \{ 0,\pm \pi\}$). This important fact
leads to the property that the interaction is not a relevant perturbation
but only a marginal one.
The CDW does not occur at arbitrarily small interaction but only beyond 
a finite threshold. If the scattering is denoted in the usual way
 a much more elaborate reasoning computing the contribution of 
 the Cooper pair and the zero sound channel
is required to yield the same result \cite{shank91,shank01}.

Exploiting the above symmetries  the number of 
 non-zero scattering amplitudes that has to be dealt with is
reduced from  $N^3$ to approximately 
$\frac{N^{3}}{32}$. 

\paragraph{Exact Results} Here we report some exact results to which
we will compare our findings.
\begin{enumerate}
\item The \emph{ground state energy} per site 
in the metallic phase $V<1$ reads \cite{yang66a}
\begin{equation}
\frac{E}{N}(V)=\frac{\cos \mu}{4}-\sin^2\mu \!\!\! 
\int\limits^\infty_{-\infty}\!\!\! \frac{dx}{
2\cosh\pi x(\cosh 2x\mu -\cos\mu)}\ ,\label{re}
\end{equation}
where $V=\cos\mu$. Eq.(\ref{re}) is dominated by its linear part 
$-(\frac{1}{\pi}+\frac{V}{2\pi})$ (cf.\ Eq.~(\ref{me})), 
which we will substract 
to focus on  the correlation part of the energy.
\item We deduce from the dispersion of the anisotropic
Heisenberg model \cite{johns73} the dispersion of the
effective fermionic model
\begin{equation}
\varepsilon_k=-\underbrace{\left(\frac{\pi}{2\mu}\right)}_{v^*_{\rm F}(V)}
\cos k\ ,\label{rd}
\end{equation}
which we expect after the CUT. In particular, we use the
value of the \emph{Fermi velocity} $v^*_{\rm F}$ as done previously 
\cite{halda81a}. This quantity is also dominated by a rather trivial linear
 term $1+\frac{2V}{\pi}$ 
(cf.~Eqs.\ (\ref{md},\ref{finitesize-disper})) which we will subtract
 in order to focus on the correlation effects.
\item
The \emph{momentum distribution} $n(k)$
in the ground state cannot be computed
directly with Bethe ansatz. But it is possible to deduce the asymptotic
behaviour for momenta close to the Fermi wave vector $k_{\rm F}$ based on
the representation of the model in terms of bosonic degrees of freedom
\cite{halda81a}. In the momentum distribution a characteristic signature
of Luttinger liquid behaviour shows up.  No finite jump  in $n(k)$ exists
at $k=\pm k_{\rm F}$ but a power law singularity with non-universal
exponent appears. This singularity is of the form \cite{halda81a}
(for more details, see e.g.~\cite{meden92,voit95})
\begin{equation}
n(k\approx k_{\rm F})\approx{\textstyle\frac{1}{2}}-C_1 \mbox{sign}(k-k_{\rm F})
|k-k_{\rm F}|^\alpha
\label{luttinger}
\end{equation}
with
\begin{eqnarray}
\label{alpha-eta}
\alpha(V)&=&\frac{1}{\eta(V)}+\frac{\eta(V)}{4}-1\label{alpha}\\
\eta(V)&=&\frac{\pi}{\pi-\arccos V}
\end{eqnarray}
and an undetermined constant $C_1$.
\end{enumerate}

\subsection{The Continuous Unitary Transformation}
In the subsection on the method \ref{method} it was explained
that the block band structure of the system is preserved under
the chosen CUT. This means that the number of quasi-particles
is altered at maximum by $4$. It must be emphasized, however, that
this does not imply that only terms made from four fermionic
operators, so-called 2-particle operators\footnote{The name stems from the
property that the effect of this operator is to re-distribute \emph{two} real 
(not quasi) particles. This is due to the conservation of the number of
real particles.},
 occur in the hamiltonian. 
Even though the initial hamiltonian (\ref{general_ham}) contains
only 0-particle (constants), 1-particle (of the form $:c^\dagger c:$)
and 2-particle terms (of the form $:c^\dagger c^\dagger c c:$) 
the transformation
may and will generate also 3-(and more) particle terms. But these terms
have to fulfill the condition that they change  the number of quasi-particles 
by no more than $\pm 4$.

 Let $T^{(j)}_{i}(\ell)$ be the normal-ordered $j$-particle term in
 $H(\ell)$ which changes the number of quasi-particles by $i$. 
So the general structure  during the transformation is
\begin{eqnarray}
H(\ell)&=&T^{(0)}_{0}(\ell)+T^{(1)}_{0}(\ell)+\nonumber\\
&&\sum_{j\geq 2}\left[T^{(j)}_{+4}+T^{(j)}_{+2}+T^{(j)}_{0}+T^{(j)}_{-2}+
T^{(j)}_{-4}\right]\!(\ell)
\label{general_st}
\end{eqnarray}
and
\begin{equation}
\label{general_ge}
\eta(\ell)=\sum_{j\geq 2}\mbox{sign}(i)T^{(j)}_i(\ell)\ .
\end{equation}
Note that the 1-particle term cannot change the number of quasi-particles
since the conservation of momentum does not allow to shift the particle
in momentum-space.

Some general analysis is possible. One can study which combinations of 
$T^{(j')}_{i'}$ and $T^{(j'')}_{i''}$ in $[\eta(\ell),H(\ell)]$ influence 
$\frac{d}{dl}T^{(j)}_{i}$. First, the changes in the
 number of quasi-particles  is additive $i=i'+i''$. 
Second, the maximum value of $j$ is
\begin{equation}
j_{\rm max} = j'+j''-1\label{estim_upper}
\end{equation}
due to the properties of the commutator between products of fermionic
operators. Third, the normal ordering yields also a lower bound for $j$
because the number of possible contractions is restricted. Contractions
in a product made from  normal-ordered factors
are possible only between the fermionic operators from \emph{different}
normal-ordered factors, see Appendix A. So we are led to \cite{heidb01}
\begin{eqnarray}
&&j_{\rm min} = \label{estim_lower} \\ 
&& \frac{1}{2}\mbox{max}
\left\{{{(j'+i')+(j''-i'')+|(j'-i')-(j''+i'')|}\atop{(j'-i')
+(j''+i'')+|(j'+i')-(j''-i'')|}}\right\}\, .
\nonumber
\end{eqnarray}

It is also possible to derive a set of abstract differential equations
for the coefficients of all 
possible terms. But its structure is quite complicated \cite{heidb01}.
 So a  solution of the complete CUTs seems to be hardly accessible and we
refrain from persuing this route further and turn to a numerical treatment
of finite-size systems  with $N$ sites. Still further approximations
are necessary because  the number of different $j$-particle processes
 grows generically as $N^{j-1}$.

\subsection{Self-Similar or Renormalizing Approximation}
We analyse the flow of all terms of the form
\begin{eqnarray}
T^{(0)}_0(\ell)&=&E(\ell)\label{t0}\\
T^{(1)}_{0}(\ell)&=&\sum_{k\in[0,2\pi)}\varepsilon_k(\ell)
:c^{\dagger}_{k}c^{\phantom\dagger}_{k}:
\label{t1}\\
T^{(2)}_{i}(\ell)&=&\sum_{{{k\in[0,\pi)}\atop{qp\in[0,2\pi)}}}
\Gamma_{kqp}(\ell)
:c^{\dagger}_{k+q}c^{\dagger}_{k-q}c^{
\phantom\dagger}_{k-p}c^{\phantom\dagger}_{k+p}:\label{t2}
\end{eqnarray}
with $i\in\{0,\pm2,\pm4\}$. Terms  involving interactions dealing 
with more than two particles $T^{(j>2)}_{i}$ are neglected.
So we proceed as follows. The sum  (\ref{general_st}) of the  terms
in Eqs.~(\ref{t0},\ref{t1},\ref{t2}) represents the hamiltonian at
a given value of $\ell$. The corresponding generator $\eta$  in 
Eq.~(\ref{general_ge}) contains  the same terms. This ansatz is
 inserted at the right hand side of
the general flow equation (\ref{general_fe}). The commutator is computed
and the result is normal-ordered (cf.\ appendix A). The 3-particle terms 
are omitted and the coefficients of the 0-, 1- and 2-particle terms are
compared. In this way, a high dimensional set of ordinary differential 
equations for the coefficients of the terms in
Eqs.~(\ref{t0},\ref{t1},\ref{t2}) is obtained.
 These equations are given in appendix B.
At $\ell = \infty$ the parts of the third term (\ref{t2}) which alter the
number of quasi-particles will have vanished so that only the $i=0$ part
remains (for illustration see Fig.~\ref{illustration_1}).
At this stage, i.e.\ at the end of the transformation,
$T^{(0)}_0$ is the ground state energy, $T^{(1)}_0$ 
the 1-particle dispersion and  $T^{(2)}_0$ represents
the interaction of two quasi-particles.
The latter does not need to be small.

Since the given structure of the initial hamiltonian $H(\ell=0)$ is preserved
 we call this approximation ``self-similar'' in the spirit of
the work by G{\l}azeck and Wilson \cite{glaze93}.
The naming ``renormalizing'' is based on three facts. First, on the technical
level the coefficients appearing in the initial hamiltonian are changed,
i.e.\ renormalized, in the course of the transformation.  Second, the 
procedure is non-perturbative since  terms are omitted not because they are
of a certain order in the initial interaction $V$
 but because of their structure being 3-(or more) particle terms.
This implies that the couplings  kept acquire infinite
orders in $V$. Third, the generator (\ref{mku_cuts},\ref{general_ge})
 which we use here leads to a smooth exponential 
cutoff $\exp(-|\Delta E| \ell)$ of the matrix elements
connecting states of different energy (energy difference
$\Delta E$) \cite{mielk98,knett00a}. 
Thus matrix elements between energetically distant states
are suppressed much more rapidly than those which are energetically very
close to each other. This is similar to what is done in Wilson's 
renormalization \cite{wilso75} where the degrees of freedom at large
energies are integrated out first.

 We illustrate the exponential cutoff
 for matrix elements connecting to the ground state. Their
asymptotic behaviour on $\ell\to\infty$ is governed by Eq.~(\ref{e-conver}).
Let us assume that the non-diagonal elements $H_{i0}$ are already very small. 
Then the diagonal energies like $E$ and $H_{ii}$ deviate from their asymptotic
values only quadratically in the non-diagonal elements as results from 
second order perturbation theory. So in leading order the diagonal elements
can be considered constant in $\ell$. Then Eq.~(\ref{e-conver})
yields $H_{i0} \propto \exp(-(H_{ii}-E)\ell)$ as stated before. 

How can the restriction to the terms in Eqs.~(\ref{t0},\ref{t1},\ref{t2})
be justified?  One argument results from considering an expansion
in the interaction $V$. The 3-particle
interactions neglected are generated by a commutator  of two 
normal-ordered 2-particle terms which are of the order of $V$ so
that they are of the order $V^2$. For our purposes it is important
to know in which order the terms kept are influenced by the neglect of
the 3-particle terms. The 3-particle terms can have an influence
on the terms kept only if they are commuted again with at least  a
2-particle term of the order $V$ or higher. Hence the neglect
of the 3-particle terms introduces
deviations only in order $V^3$. This holds for the 1- and 2-particle
terms (\ref{t1},\ref{t2}). The 0-particle term, which becomes at $\ell=\infty$
the ground state energy, is not changed by the commutator of a 2-particle
and a 3-particle term since the generated terms cannot be contracted completely
as stated by Eq.~(\ref{estim_lower}). So the deviation  in the
0-particle term engendered by the
neglect of the 3-particle terms is at worst of order $V^4$.
Thus the approximation can be justified for low values of $V$.

Another argument, which is more general than a power counting 
in the interaction, comes from the structure of the terms neglected. 
Let us focus on the 3-particle terms. Since they may change the
number of quasi-particles at most by $4$ they contain at least one
annihilator of a quasi-particle. Since they are normal-ordered this
annihilator appears rightmost. Hence such a term is active only if
applied to a state which contains already some excitations. 
Hence the approximation chosen is justified if the system can be considered
a \emph{dilute} gas of quasi-particles irrespective of the strength
of the interaction between the quasi-particles. 
In the course of the transformation  virtual processes
creating and annihilating quasi-particles are more and more suppressed so that
the average concentration of quasi-particles decreases gradually. Hence towards
the end of the transformation the neglect of 3-(and more) particle terms
is well justified. On the other hand, the 3-(and more) particle terms
are not present in the initial hamiltonian (\ref{general_ham}) so that their
neglect is well justified during the first phase of the transformation as well.
Only during the intermediate phase there is no general control of the
quality of the approximation. Here one has to focus on the particular system
 under study. To assess the quality of the approximation
one may either compare to otherwise available data on the system 
(external quality control) or one may compute some of the terms neglect 
in order estimate how large they are (internal or self-consistent quality
control). In the present work we will use the first method and 
assess the quality of our findings by comparing them to the known
exact results.
Summarizing, the approximation is well justified if the system can be 
considered a dilute gas of quasi-particles independent of the actual
interaction between the quasi-particles.

Finally, we  wish to 
point out an analogy between our approach and a more standard
diagrammatic one. By restricting ourselves to the terms in 
Eqs.~(\ref{t0},\ref{t1},\ref{t2}) we are dealing with a 2-particle irreducible
vertex function which is renormalized continuously. The 
renormalization equation (see appendix B) is bilinear in all coefficients.
The terms contributing to $\partial_\ell \Gamma$, which are
bilinear in the vertex function $\Gamma$,
result from the commutation and a single contraction of two 2-particle 
irreducible vertex functions.
In this sense our procedure bears similarities to a 1-loop renormalization 
or to the summation of all parquet diagrams \cite{solyo79}.
The main difference  is that our approach keeps the problem local in time
along the renormalization flow because the transformation is unitary.
Hence not frequency dependence enters. We consider this a major
advantage of the CUT approach for numerical application since much less
bookkeeping is required since no frequency dependence has to be traced.

\section{Numerical Results}
We turn now to the numerical solution of the differential 
equations set up in appendix B. The momentum dependence of all
functions is discretized in an equidistant mesh. The positions of
the points of this mesh are chosen such that no $k$-point lies 
precisely at $k=\pm k_{\rm F} = \pm \pi/2$. For $N$ divisible by 4,
this corresponds to antiperiodic boundary conditions. In this way 
unnecessary degeneracies are avoided. For large system size
the influence of the boundary conditions  becomes increasingly
unimportant. 

For the system sizes ($N\approx 50$) 
that we will be looking at there are about 50.000
coupling constants which are traced in the set of differential equation.
Luckily, the differential equations are not very sensitive since
they describe the convergence to a static fix point. 
The numerics is done by a Runge-Kutta algorithm with 
adaptive stepsize control. This algorithm is robust but quite
 laborious because the  differential equations have to be evaluated 
six times for each step. Rigorously, the fix point is reached at
 $\ell=\infty$. In practise, we stop the flow when the relative change
of  $E$ and $\varepsilon_k$ falls below  $10^{-6}$.

\subsection{Illustration of the CUTs}
\begin{figure*}
\centering
\mbox{
\epsfig{height=5cm,figure=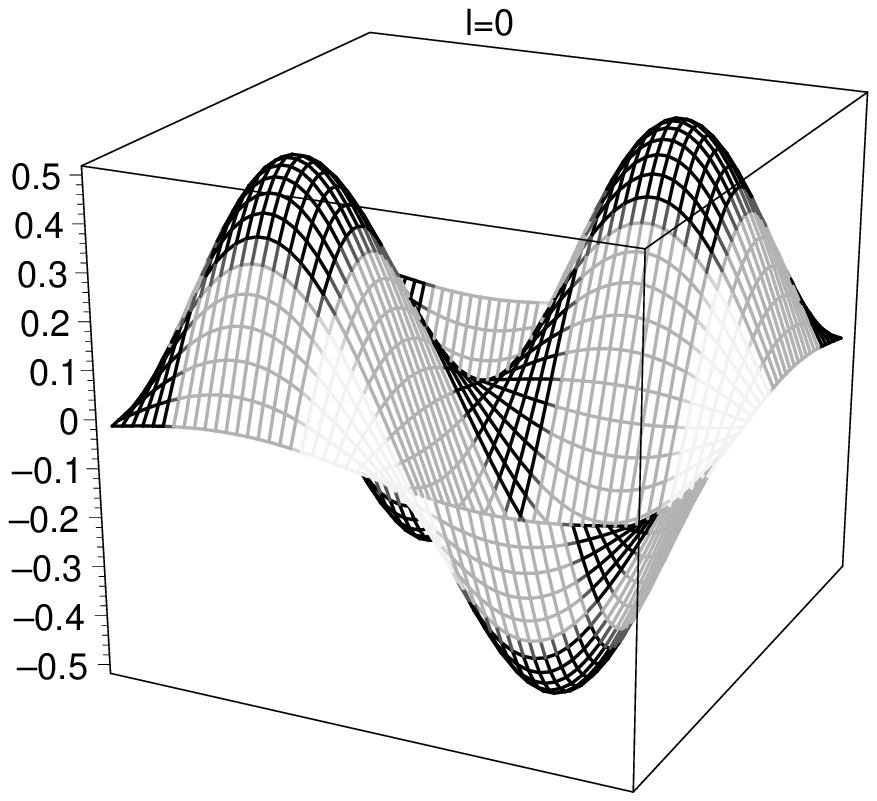}
\epsfig{height=5cm,figure=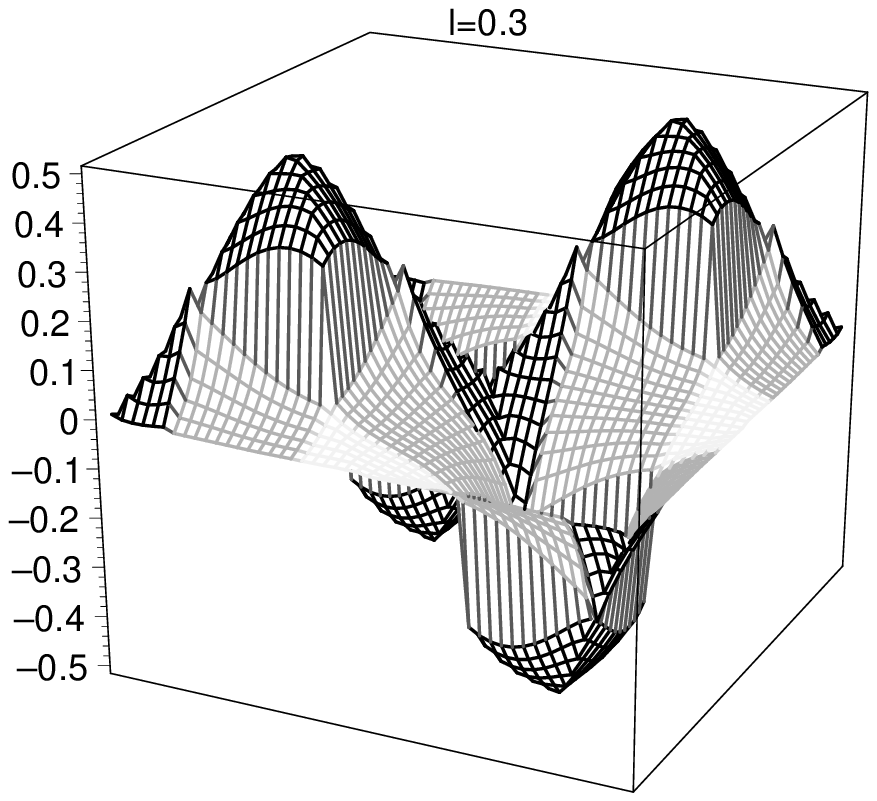}
\epsfig{height=5cm,figure=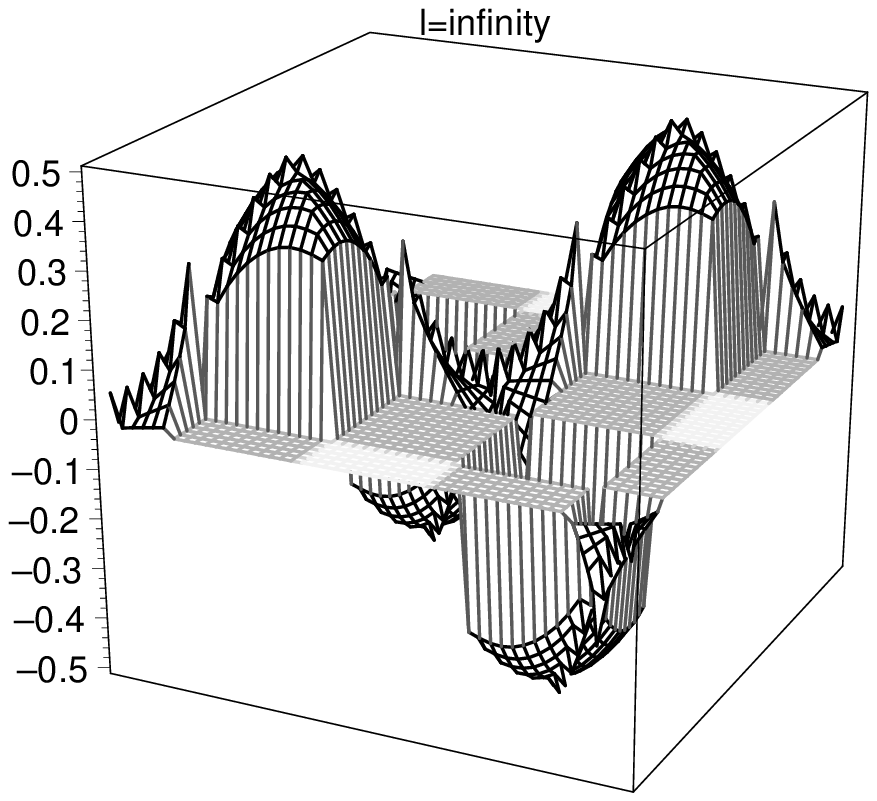}
}
\caption{Sections of the vertex function  $\Gamma_{kqp}$
at $k=\frac{\pi}{4}$ as function of
 $q,p\in\{-\pi,\pi\}$ at three values of $\ell$ ($0,0.3$
and $\infty$) for $V=0.5$ and $N=48$ sites.
 The lightest parts are those that change the number of 
quasi-particles by $\pm 4$, the
 gray parts change them by $\pm 2$, the black parts leave it unchanged.
 The quasi-particle-number changing amplitudes are transformed to zero.}
\label{illustration_1}
%\end{figure*}
%\begin{figure*}
\centering
\mbox{\hspace{-1cm}
\subfigure[Symmetric replacement]{
\parbox{7.2cm}{
\center
\epsfig{width=30mm,figure=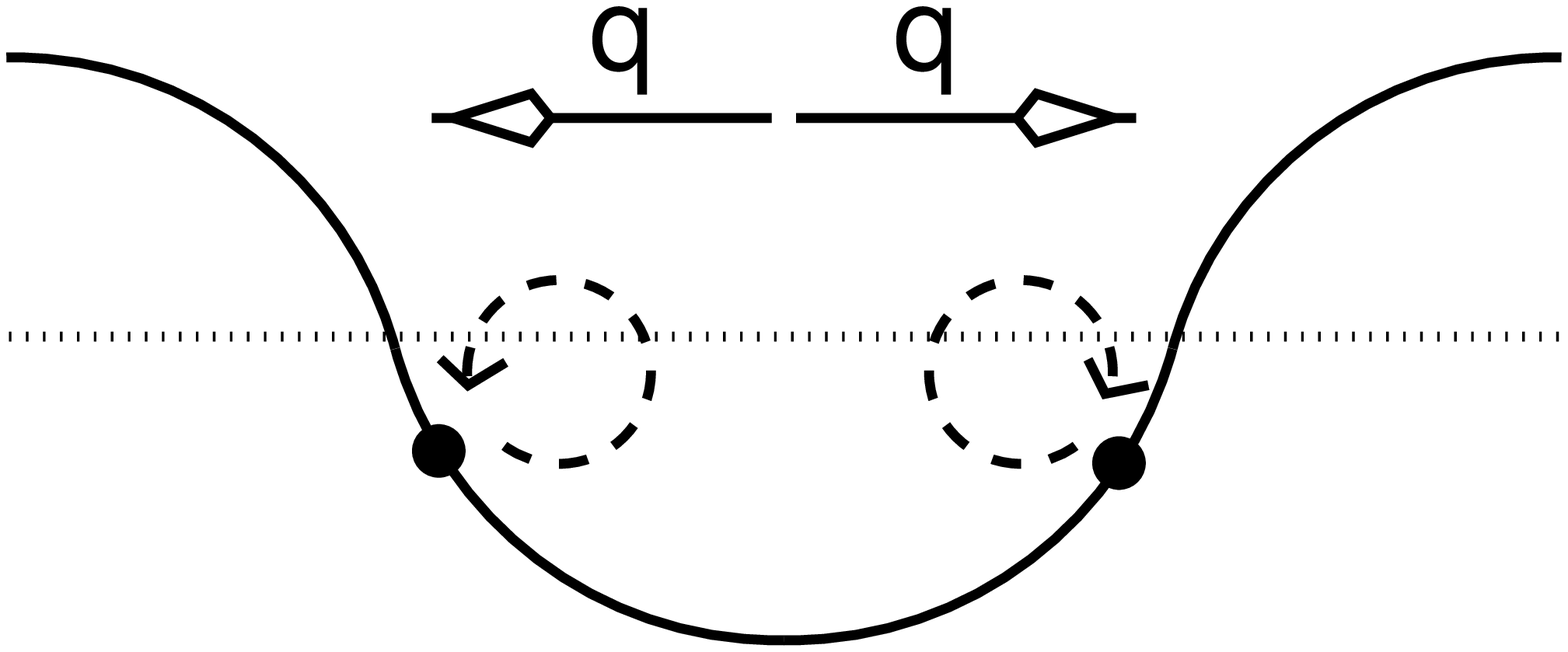}
\epsfig{width=7.2cm,figure=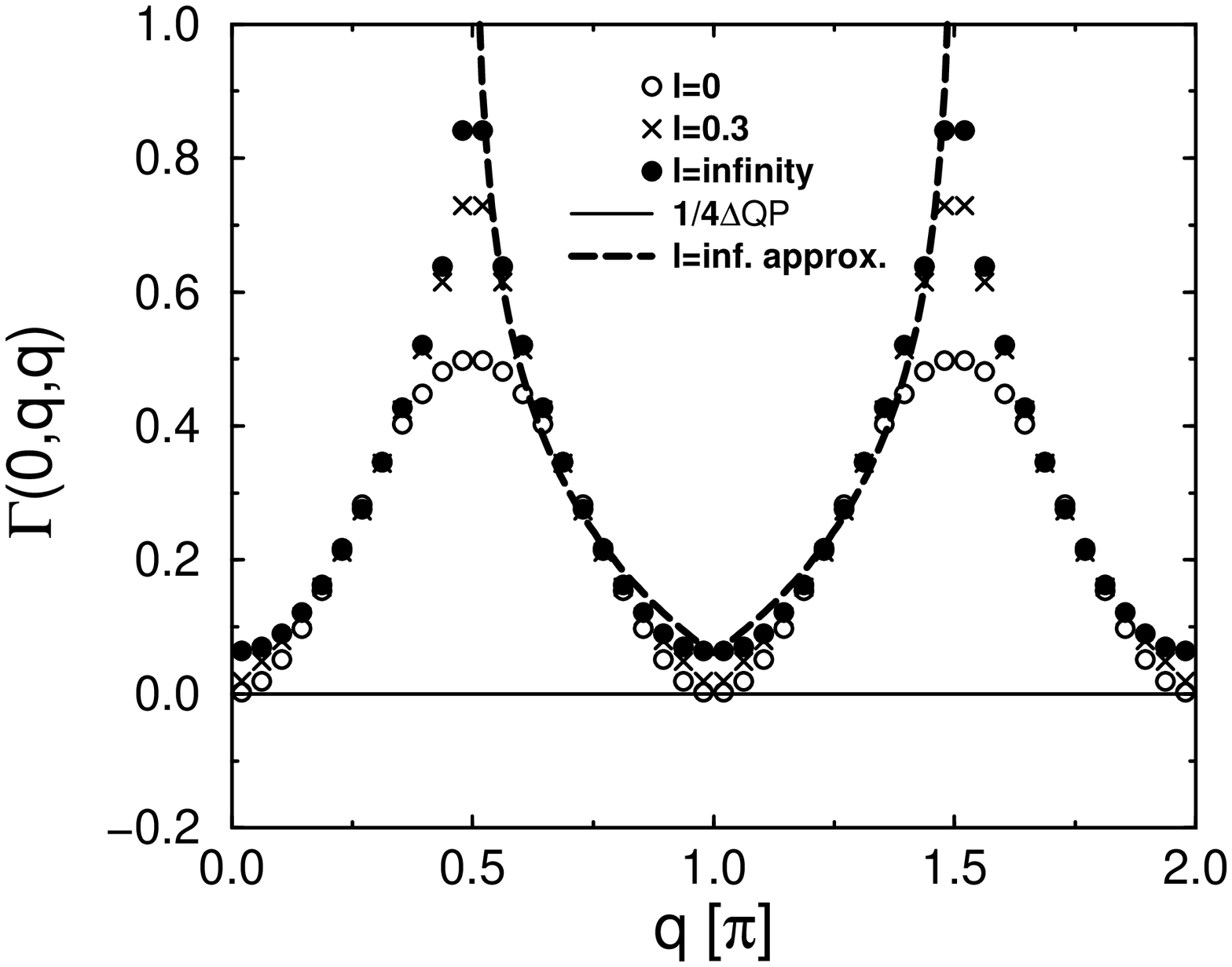}
\vspace{-0.75cm}
\label{illustration_2a}
}}
\hspace{1cm}
\subfigure[Symmetric scattering of two particles 
creating (annihilating) four quasi-particles]{
\parbox{7.2cm}{
\center
\epsfig{width=30mm,figure=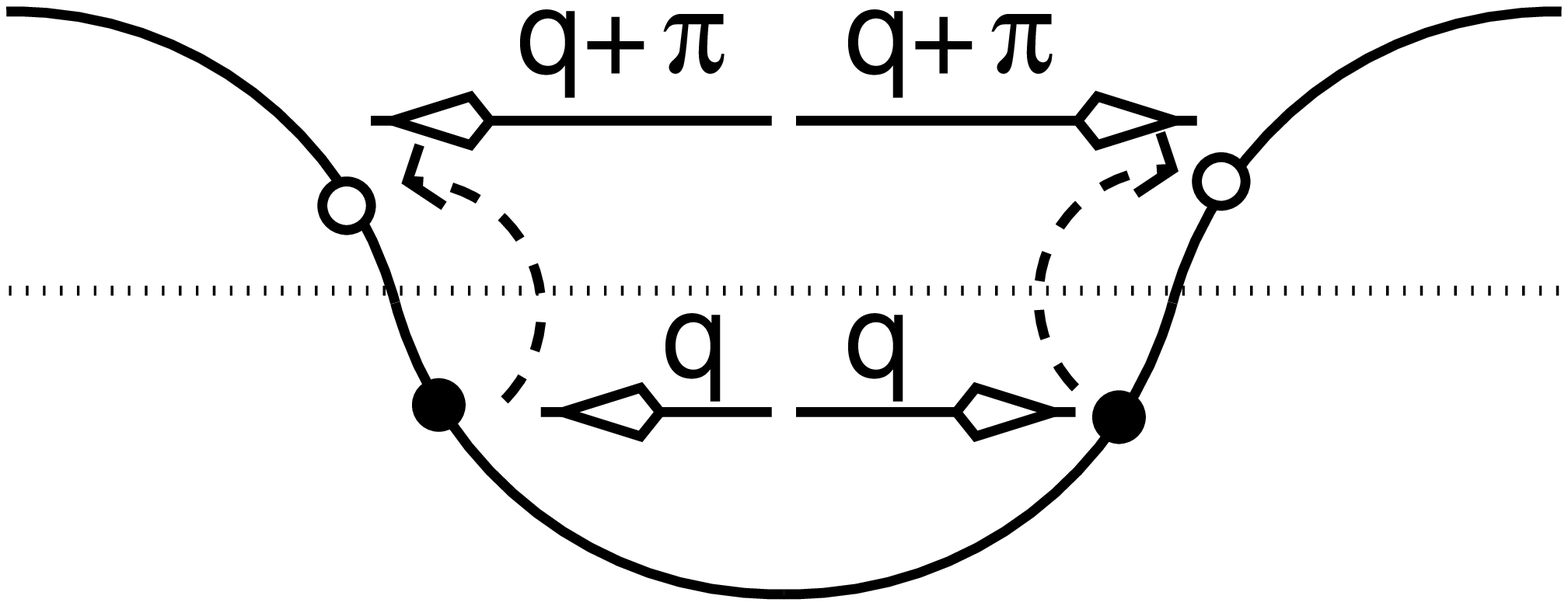}
\epsfig{width=7.2cm,figure=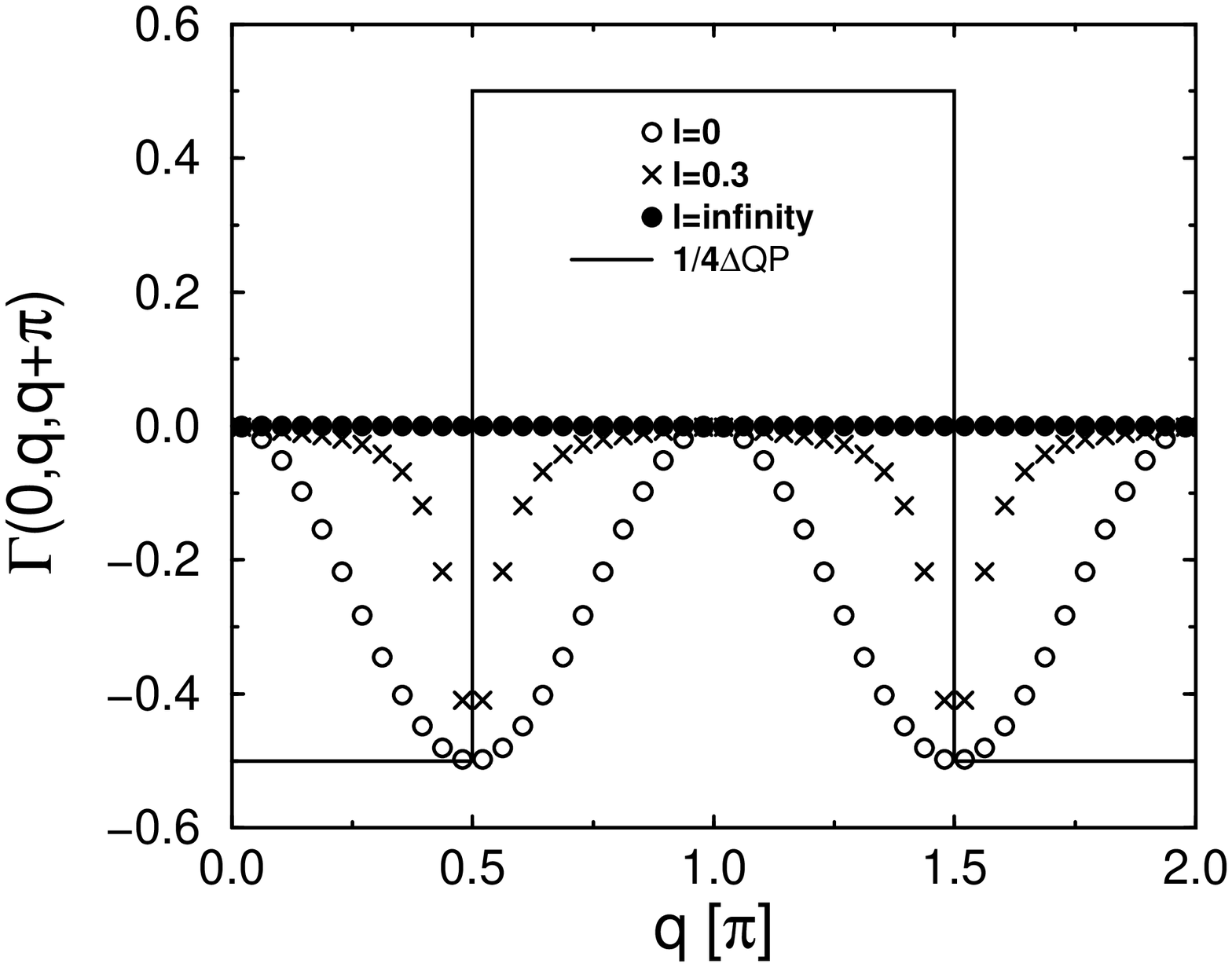}
\vspace{-0.75cm}
\label{illustration_2b}
}}
}

\vspace{-6mm}

\mbox{\hspace{-1cm}
\subfigure[Changing the average momentum $k$ 
keeping one  scattering process unchanged]{
\parbox{7.2cm}{
\center
\epsfig{width=30mm,figure=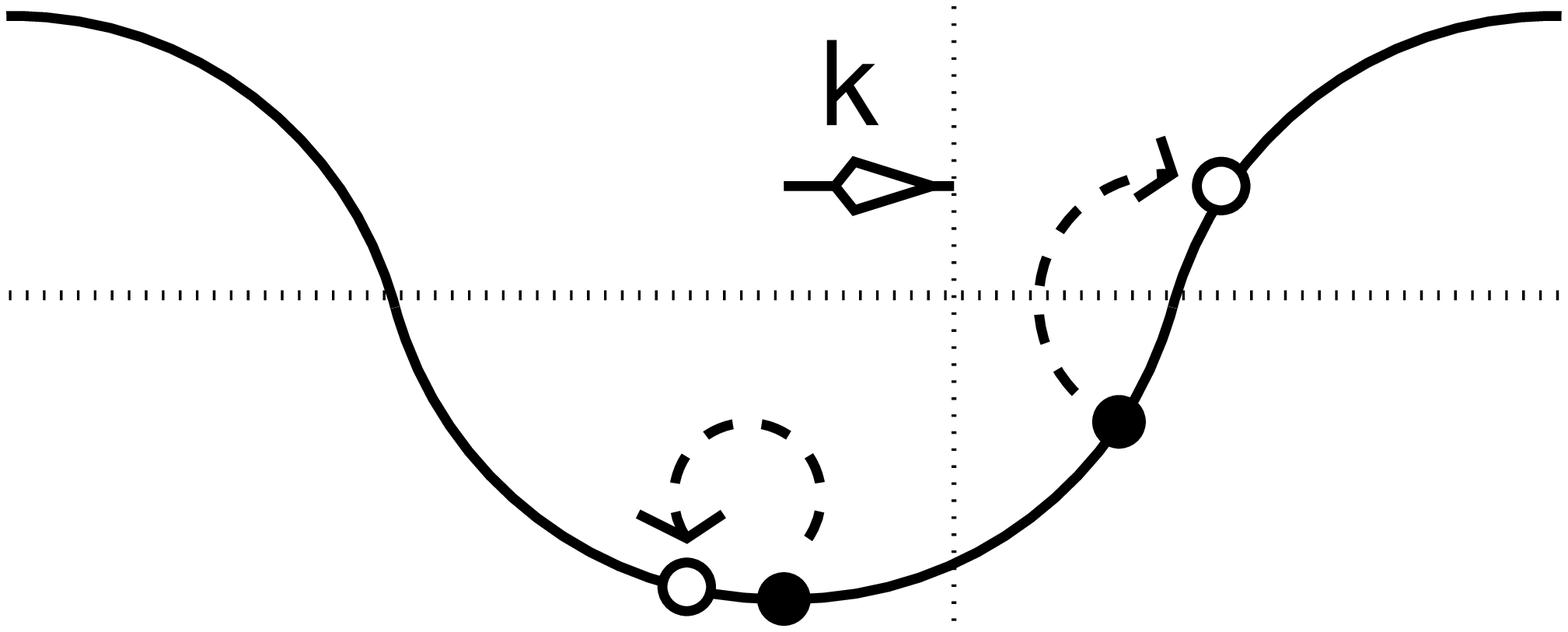}
\epsfig{width=7.2cm,figure=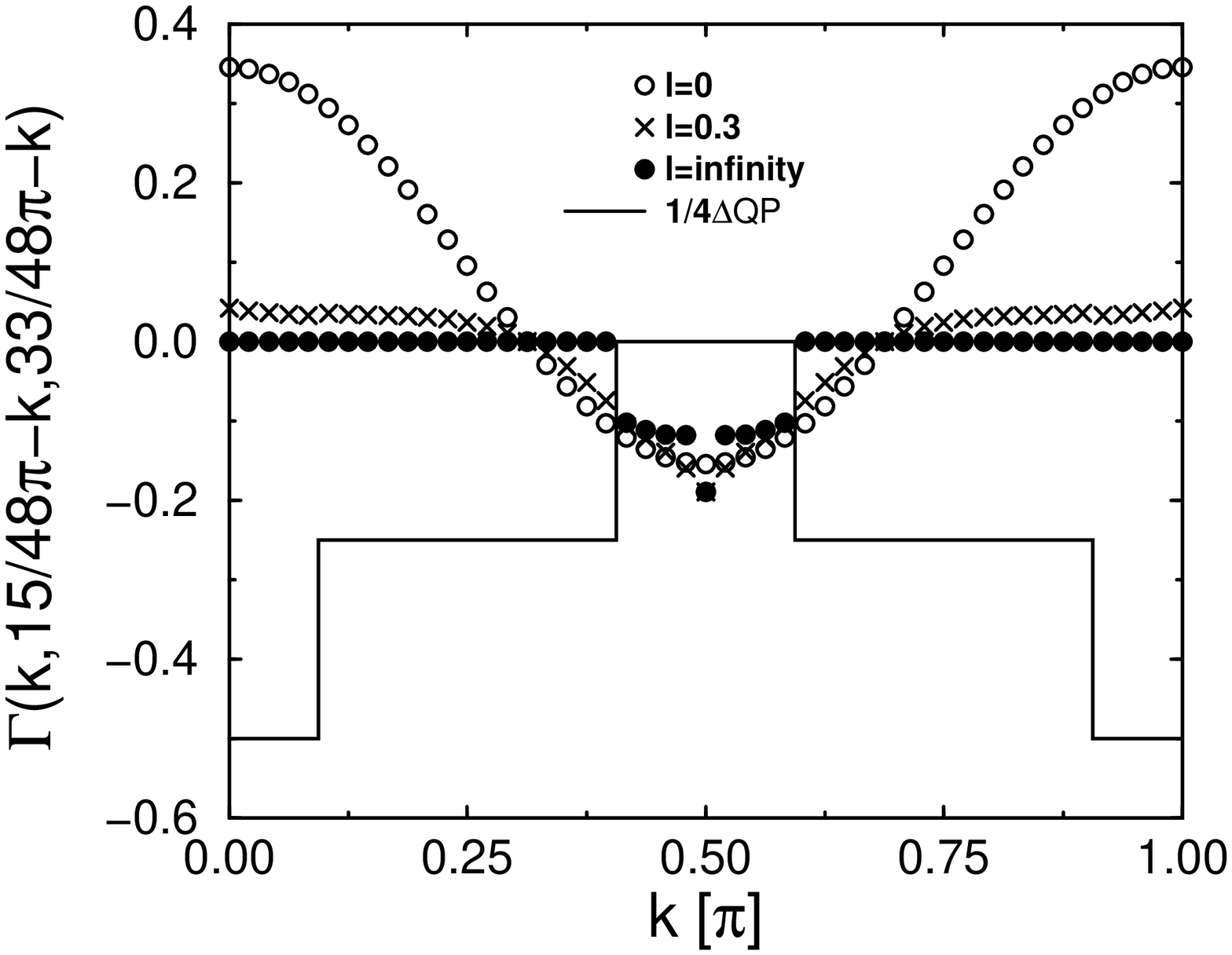}
\vspace{-0.75cm}
\label{illustration_2c}
}}
\hspace{1cm}
\subfigure[From forward-scattering ($q\approx 0$) to umklapp-scattering
($q\approx \pi$) and back ($q\approx 2\pi$)]{
\parbox{7.2cm}{
\center
\epsfig{width=30mm,figure=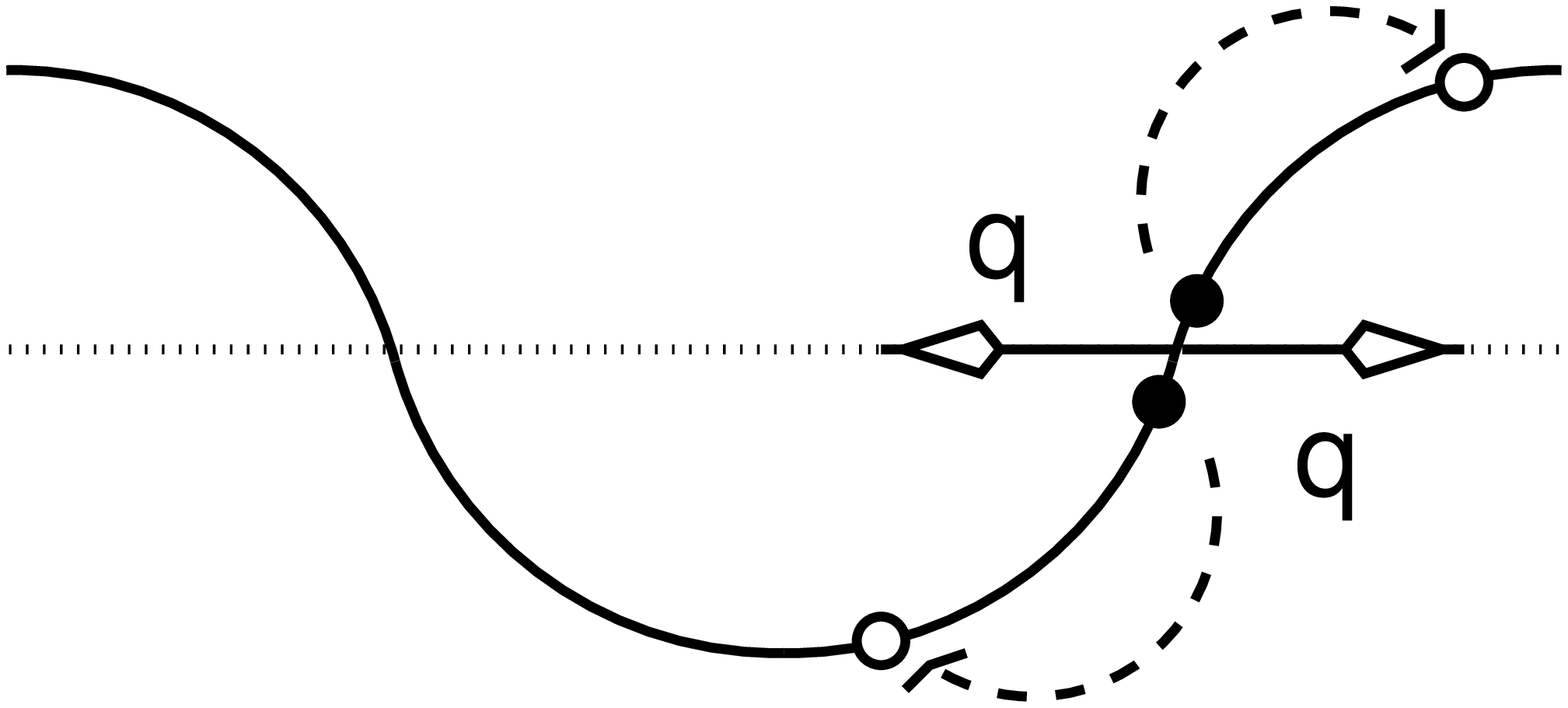}
\epsfig{width=7.2cm,figure=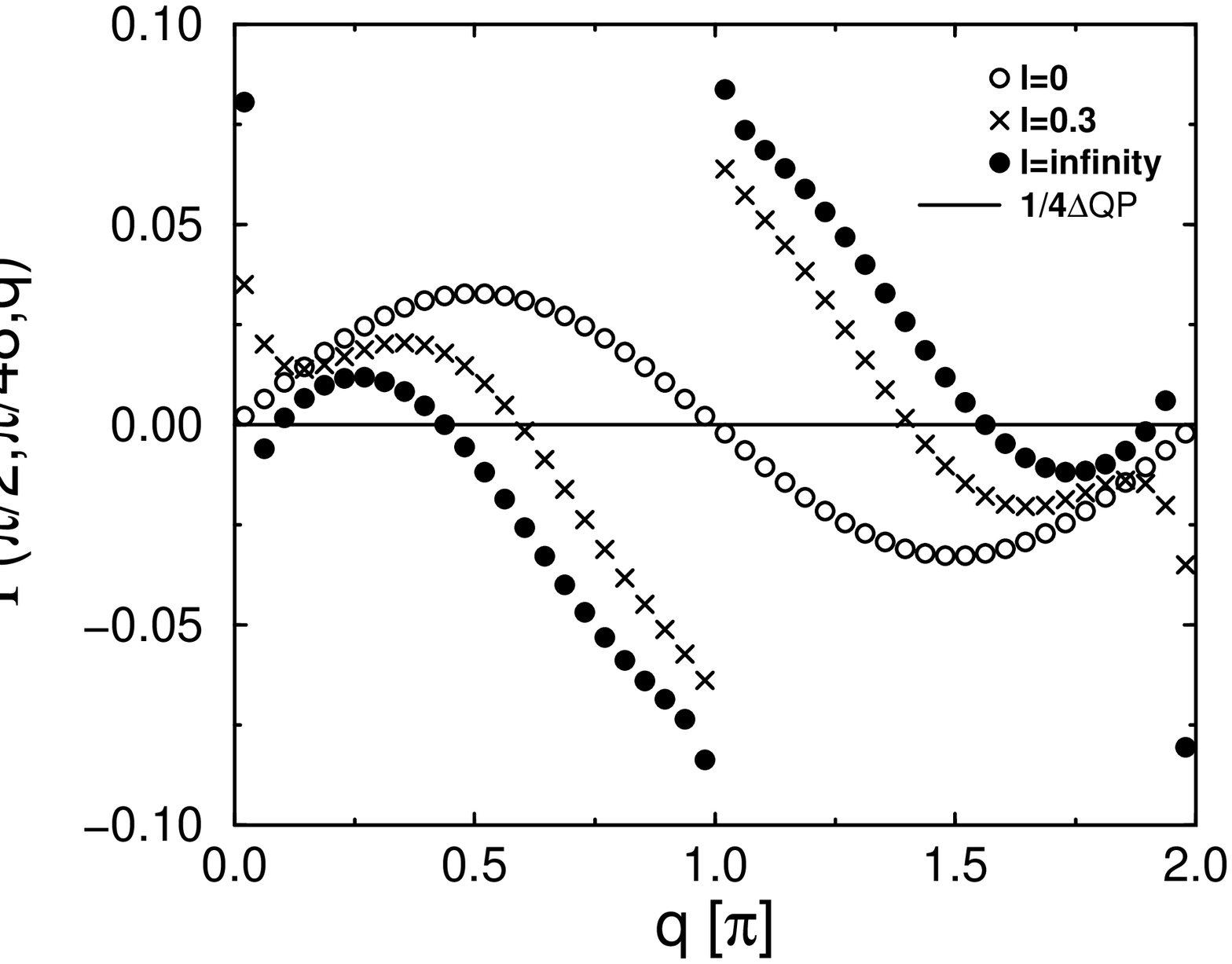}
\vspace{-0.75cm}
\label{illustration_2d}
}}
}
\vspace{-2mm}
\caption{Sections of  the 
vertex-function $\Gamma_{kqp}$  as in
$\Gamma_{kqp} :c^{\dagger}_{k+q}c^{\dagger}_{k-q}
c^{\protect\phantom\dagger}_{k-p} c^{\protect\phantom\dagger}_{k+p}:$. 
The pictograms illustrate the scattering processes. 
They show the dispersion (solid line), filled dots for the fermions 
to be annihilated and open dots for the holes the fermions
are put. The arrows depict how the processes change on varying the
momentum. $\Delta$QP denotes the change of
the  number of quasi-particles induced by the scattering process. Note that
$\Delta$QP $\neq 0$ implies that the amplitude vanishes for 
$\ell=\infty$.}
\label{illustration_2}
\end{figure*}

Fig.~\ref{illustration_1} illustrates what the continuous unitary 
transformations are doing. The parts of the 
vertex-function that are associated with processes that change the number of 
quasi-particles are transformed away as $\ell$ goes to $\infty$
 while the other part is renormalized. 
At the end of the transformation there are 
discontinuities at the borderlines of the different parts and some
singular kinks if the scattering processes occur at the Fermi points
or if $p=q$ holds.

A more quantitative
 insight is given by Fig.~\ref{illustration_2} where one-dimensional 
sections of the vertex-function $\Gamma_{kqp}$ are shown.
\begin{itemize}
\item 
Fig.~\ref{illustration_2a} depicts  $\Gamma_{0qq}$ for various
 values of $\ell$. These amplitudes are  related to processes 
where two fermions at $q$ and $-q$ are first annihilated and then created
again. The shape of the function close to $|q|=\pi/2$ can be fitted by 
$\Gamma_{0qq}\approx -0.007q^2-0.266\ln||q|/\pi-0.5|-0.125$.
So a logarithmic singularity at $\frac{\pi}{2}$ can be presumed.
\vspace{+2mm}
\item 
In Fig.~\ref{illustration_2b} the amplitudes of processes are shown
where two  fermions are taken from $q$ and $-q$ and put at
 $\pi-q$ and $-\pi-q$. All these processes change the quasi-particle 
number and, hence, have to vanish for $\ell\to\infty$. The processes
 near the Fermi wave vector $q\approx \frac{\pi}{2}$ are 
decreasing very slowly.
\vspace{+2mm}
\item 
For the plots in Fig.~\ref{illustration_2c} we fix one shift of a fermion
from below the Fermi level to an energy above it. The shift of the
other fermion is varied.  Here, different changes
 in the number of quasi-particles are possible. There are regions 
where the amplitudes vanish and others where they are only renormalized.
The process at $k=\frac{\pi}{2}$ is apparently different. It corresponds
to the exchange of the two particles. At present, we cannot judge whether
this difference will be relevant in the thermodynamic limit. It is an
interesting question whether it retains a finite weight for $N\to \infty$.
\vspace{+2mm}
\item 
In Fig.~\ref{illustration_2d} we show the evolution of a scattering
event from pure exchange at $q=0$ at the right Fermi point over
forward-scattering for small positive values of $q$ to 
umklapp-scattering at $q=\pi$. One realizes that the scattering
amplitudes at $q=0$ and at $q=\pi$ are particularly enhanced
by the renormalization. It seems totally inappropriate to approximate
the scattering amplitudes as being constant for small momentum ($q\approx 0$)
or for large momentum ($q\approx\pi$) because the points $q=0$ and
$q=\pi$ appear to be decisively different from the scattering in their
vicinity. This observation makes the question whether a 
 continuum description is quantitatively applicable an interesting issue
for future studies.
\end{itemize}

\subsection{Convergence}
\begin{figure}
\center
\epsfig{width=7cm,figure=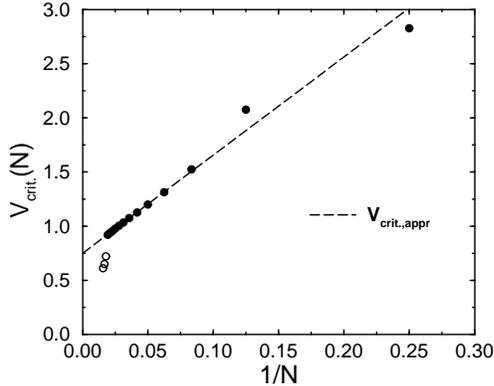}
\caption{Interaction values above which the approximation breaks down 
because of loss of convergence. Filled symbols stand for numerically
stable runs, open symbols ($N\geq 52$) for runs where numerical 
inaccuracies spoil the convergence. The dashed line extrapolates 
the numerically reliable results (see Eq.~\ref{vcrit}).}
\end{figure}

General statements on the convergence of the CUT approach chosen
are possible \cite{mielk98,knett00a}. But they do not ensure
that the method works for macroscopically large systems. 
Additionally, the unavoidable use of approximations may lead to
the break down of convergence. So we investigated empirically
up to which value of the interaction $V_{\rm crit}$ the CUT
works. This value depends on the system size $N$ or the density of the
discretization mesh, respectively. 
If $V$ is larger than $V_{{\rm crit.}}(N)$ all couplings are 
diverging at a certain $\ell$ and the solution of the differential 
equations cannot be continued asymptotically to $\infty$. 
Furthermore, $V_{{\rm crit.}}(N)$ is 
decreasing as $N$ is growing. In addition, numerical problems occur at system 
sizes $N>52$  and $V$ close to $V_{{\rm crit.}}(N)$
due to accumulated inaccuracies. This can be seen
 in the unsystematical behaviour of 
$V_{{\rm crit.}}(N)$ above $N=52$ and in the unphysical 
shape of e.g.\
 the dispersion for $V$ close to $V_{{\rm crit.}}(N)$ in
 Fig.~\ref{dispersion_film}.
On the other hand, the decrease of $V_{{\rm crit.}}(N)$ is very 
systematic as long as $N\leq 52$ holds. It can 
 be  approximated very well  by a linear dependence
\begin{equation}
\label{vcrit}
V_{{\rm crit.}}(N)\approx {9.06/N}+0.75 \ ,
\end{equation}
which leads to an extrapolated value in the thermodynamic limit 
$N\to\infty$ of $V_{{\rm crit.}}(\infty)
\approx 0.75$. We conclude that it is possible to study the 
metallic phase of the system and that the conclusions drawn from
the finite-size calculations are also relevant for the thermodynamic
limit. A description of the transition to the
insultating phase is presently not possible.

\paragraph{The Case $N=4$}

\noindent
Some insight on the cause for the loss of convergence can be obtained
in the simple case of four points in momentum space, which is
analytically solvable with and without approximation. The analytic solution
is  simple because the quasi-particle vacuum $|\psi_0\rangle$ is  connected
by the interaction only to the state
where both fermions are excited $|\psi_{\scriptsize ex.}\rangle$. Thus
one has  to diagonalize a $2\times2$ matrix which can be done 
easily directly or using the CUT.

Our approximation, however, is dealing with operators and not with
matrix elements. The state $|\psi_{\scriptsize ex.}\rangle$ involves four 
quasi-particles - two holes and two excited particles. In the approximation
3- and 4-particle terms are neglected so that the energy of the excited
state is  incorrect. Indeed, the gap between the ground state and the
excited state is under-estimated\footnote{But the deviation occurs only in
fourth order in $V$.}. 
It even closes for $V\geq \sqrt{8}$ 
and the couplings diverge before reaching $\ell=\infty$. Thereby, it is
shown that the approximation may spoil the applicability of the approach.

\subsection{Ground State Energy per Site}
\begin{figure}
\center
\epsfig{width=7cm,
figure=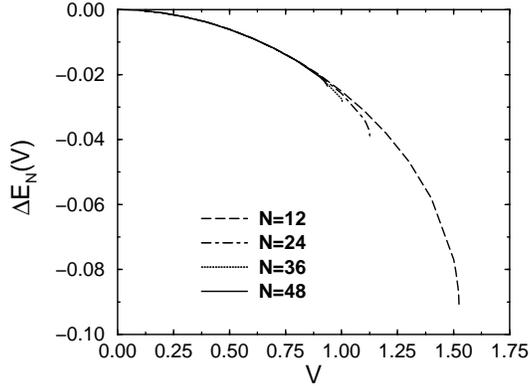}
\label{groundstate_1}
\caption{CUT results for the correlation part, i.e.~beyond
linear order, of the ground state energy 
per site for various system sizes $N$.}
\end{figure}
\begin{figure}
\center
\epsfig{width=7cm,figure=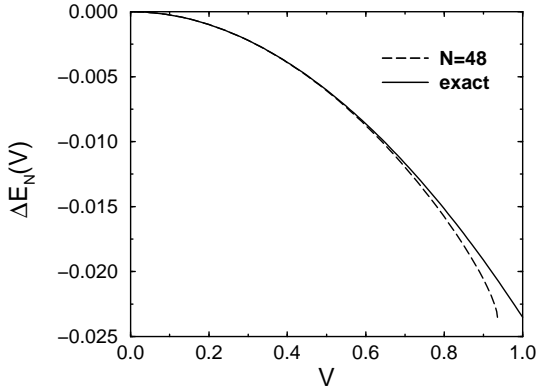}
\label{groundstate_2}
\caption{Comparison of the CUT results ($N=48$)
 for the correlation part of the ground state energy  per site
with the exact thermodynamic result.}
\end{figure}
In Fig.~\ref{groundstate_1} we plot the shape of the non-linear lowering of the
 ground state energy, i.e.\ the correlation part of the
ground state energy, for various system sizes $N$ as obtained by CUT.
The ground state energy diverges to negative values close to
$V_{{\rm crit.}}(N)$. For $V$ smaller than the extrapolated value 
$V_{{\rm crit.}}(\infty)$ all systems show nearly the same
dependence on the interaction $V$.

The case $N=48$ is compared to the exactly known 
thermodynamic result \cite{yang66a,yang66b} in Fig.~\ref{groundstate_2}. 
The quantitative results are very close to each
 other for quite a large region of $V$. Even for $V=0.5$ the relative 
difference is less than 1\%. Only as $V$ approaches 
$V_{{\rm crit.}}(48)=0.9361$ the approximate ground state energy is diverging 
very fast to $-\infty$.

\subsection{Dispersion and Renormalized Fermi Velocity}
\begin{figure}
\center
\epsfig{width=7cm,figure=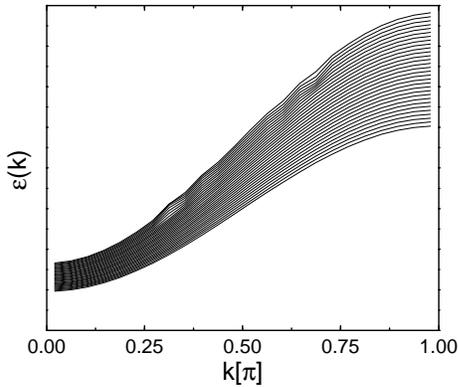}
\caption{Renormalized dispersions from the CUT calculation for
 $N=48$. The results for $V=0.03,0.06, 0.09\ldots,0.9$ are shifted with 
respect to each other in order to yield a three-dimensional view on
the evolution of the dispersion.}
\label{dispersion_film}
\end{figure}
At the end of the transformation, i.e.\ at $\ell=\infty$,
when the effective hamiltonian has reached block diagonality
only scattering processes are left which leave the number of quasi-particles 
unchanged. Then $\varepsilon_k$ is the renormalized 
1-particle dispersion. This means that in the 
 effective model \emph{after} the transformation
it is possible to add a single 
quasi-particle (hole or particle) to the ground state such that
the resulting state is an exact eigen state because there is no other
quasi-particle to interact with.

Note that this statement is \emph{not} in contradiction with the
widely known fact that the single-particle propagator $G(k,\omega)$
 of a Luttinger
liquid does not display quasi-particle peaks 
\cite{meden92,schon93,voit93a,voit95} because the 
single-particle propagator $G(k,\omega)$ refers to adding or
taking out a fermion \emph{before} any transformation.
It is an interesting issue, yet beyond the scope of the present work,
to apply the CUT (\ref{operator_fe}) 
to the creation and annihilation operators in order to recover the
usual Luttinger liquid result in the framework of the CUT renormalization.

Fig.~\ref{dispersion_film} shows the evolution of the dispersion 
for $N=48$ on increasing $V$. The dispersion behaves like a 
cosine-function with a renormalized Fermi velocity as expected from the exact 
result. For $V$ close to $V_{{\rm crit.}}(48)$ (at about $V=0.7$)
 there are kinks emerging. We reckon that these kinks represent spurious
features induced by accumulated numerical inaccuracies for the same reasons
for which the convergence is hampered for large system sizes.

\paragraph{Renormalized Fermi Velocity}
\begin{figure}
\center
\epsfig{width=7cm,figure=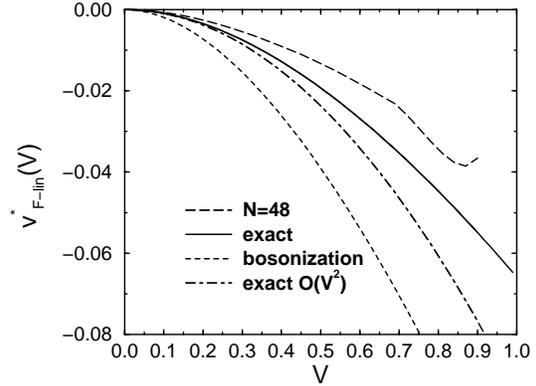}
\caption{Comparison of the non-linear correlation part of the
Fermi velocity as obtained for $N=48$ by the CUT with the exact
thermodynamic result, the exact quadratic order $V^2$ and the
result obtained from bosonization.}
\label{v_F}
\end{figure}
By fitting the function\\ $-v^*_{\rm F}\cos(k)$ to
the calculated dispersions we obtain results for the renormalized
Fermi velocity  $v^*_{\rm F}$ as function of the interaction $V$.
These values are dominated by the  linear Fock term $1+\frac{2V}{\pi}$ 
or its finite-size equivalent in (\ref{finitesize-disper}), respectively. 
In order to yield a better resolution of the influence of correlations
 we substract the constant and linear terms. The remainder $v^*_{\rm F-lin}$
is plotted in  Fig.~\ref{v_F}. It is compared to the exact result
\cite{johns73}, the exact quadratic term $V^2$ and the result obtained
from bosonization (cf.\ appendix C). The shape of the curves is 
basically the same. The CUT result is too small (in modulus) compared
to the exact results. For small $V$ this is mainly due to a finite-size 
effect as comes out from an extrapolation $N\to\infty$. Again, the CUT
approach is less reliable close to the critical interaction
value $V_{\rm crit}(N)$.

Note that the bosonization results fits less well to the exact result
than does the CUT result. This is due to the fact that in bosonization
only the processes infinitely close to the Fermi points are considered.
The deviation between the dash-dotted second order curve from the
exact results reveals that higher order terms are  important as well.
They are partly captured by the CUT procedure.

\subsection{Momentum Distribution}
\begin{figure}
\center
\epsfig{width=7cm,figure=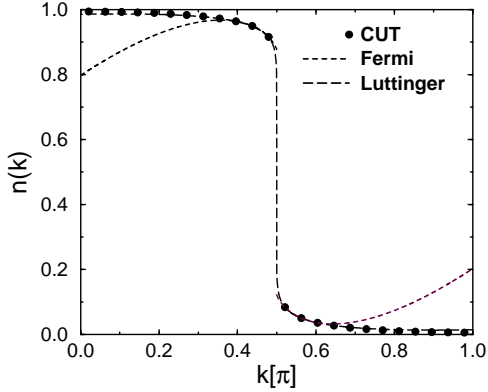}
\caption{Momentum distribution as obtained by CUT
for $V=0.6$ and 48 sites. Lines are interpolations to the
three points closest to the Fermi wave vector
assuming  Luttinger (\ref{lutti}) or Fermi behaviour (\ref{fermi}),
 respectively.}
\label{distribution}
\end{figure}
\begin{figure}
\center
\epsfig{width=7cm,figure=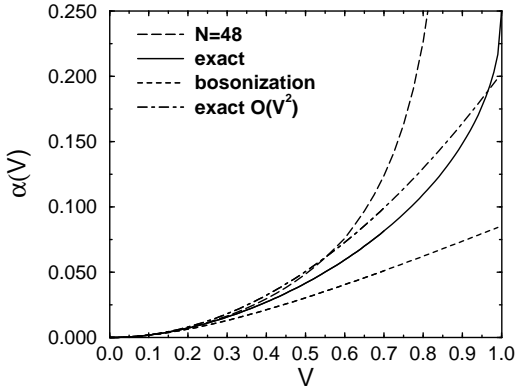}
\caption{Luttinger exponent $\alpha(V)$ as obtained from least-square
fits of the  CUT momentum distributions.}
\label{alpha_pic}
\end{figure}
Last but not least
 we analyze the momentum distribution $n(k)$ in order  to show that 
 Luttinger liquid behaviour is retrieved. The standard approach to do so
would be to transform the operator  $c^{\dagger}_kc^{\phantom\dagger}_k$
according to Eq.~(\ref{operator_fe}) \cite{wegne94,hofst00}.
Here, however, we use a simpler approach suitable for static expectation
values and correlations. The value $E_\infty=E(\ell=\infty)$
is the ground state energy within our approximation because the block diagonal
effective hamiltonian $H(\ell=\infty)$ does not contain  processes
exciting the vacuum any more. First order perturbation theory
shows straightforwardly that functional derivation of $E_\infty$
yields the expectation value $n(k)$
\begin{equation}
\label{funderiv}
\frac{dE_\infty}{d\varepsilon_k}=
\left\langle\frac{dH}{d\varepsilon_k}\right\rangle = 
\langle:c^{\dagger}_kc^{\phantom\dagger}_k:\rangle =: n(k) \ .
\end{equation}
In the numerical treatment the functional derivation can be easily
realized by approximating the ratio of infinitesimal differences by
the ratio of small finite differences. 
Hence no serious extension of the algorithm is needed to compute the momentum 
distribution. For relative variations of $\varepsilon_k$ 
 around $10^{-3}$ the ground state energy $E_\infty$ is 
linear in these variations within an accuracy of about $10^{-5}$.
So $n(k)$ can be determined to this accuracy.
 In Fig.~\ref{distribution} the 
momentum distribution (symbols)
 for $N=48$ and $V=0.6$ obtained in this way is depicted.

Due to the discretization $n(k)$ can be computed only for a finite
set of points. At first sight, a jump seems to dominate at the
Fermi wave vector $\pi/2$. But a \emph{discretized} power law distribution 
displays also a jump -- in particular if the exponent is small.
In order to understand the nature of the distribution 
a quantitative analysis is required. Hence we fit the distribution 
obtained to two functions, one being appropriate for
describing a Luttinger liquid \cite{halda81a,voit95}
\begin{equation}
 \left|n(k)-\textstyle\frac{1}{2}\right| \approx 
C_1 (\Delta k)^\alpha +C_2 \Delta k
\label{lutti}
\end{equation}
the other being appropriate for describing a Fermi liquid
 \cite{noack99}
\begin{equation}
\left|n(k)-\textstyle\frac{1}{2}\right| \approx
\frac{1+Z_{k_{\rm F}}}{2}+C_1\ln(\Delta k)
\Delta k+C_2 \Delta k
\label{fermi}
\end{equation}
with $\Delta k:= |k-k_{\rm F}|$.
For both possibilities  three free parameters 
($\alpha$, $C_1$, $C_2$ or 
 $Z_{k_{\rm F}}$, $C_1$, $C_2$, respectively) are determined.
In Fig.~\ref{distribution} the parameters are fixed to interpolate
the three points closest to the Fermi wave vector. Another way
is to perform a least-square fit; the resulting curves are shown
in Fig.~3 in Ref.~\cite{heidb02a}. In both analyses, the qualitative 
result is the same. The Luttinger fit describes our data much better
than the Fermi fit. We conclude that our data  describes rather
a power law behaviour than a jump. We cannot exclude, however,
a behaviour comprising a power law behaviour \emph{and} a jump.
But there is no reason to believe that such a behaviour should occur.

Due to the small exponents $\alpha$ occurring and the restricted
system sizes the power law 
cannot be distinguished reliably from a logarithmic behaviour or from
a function of some logarithm in $|k-k_{\rm F}|$. Note, however,
that the simple logarithm found in Ref.~\cite{wegne94} is not
likely to occur since we do not transform the observable in 
leading order only. Infinite orders of the interaction $V$ contribute
to the ground state energy and hence to the derivative (\ref{funderiv}).

In order to push the analysis one step further we compare the
exponent $\alpha$ resulting from the fits to the exact one. 
The points in the vicinity
of the Fermi wave vector are more influenced by finite size effects
\cite{schon96}. Eventually, we choose the exponents coming
from least-square fits. In Fig.~\ref{alpha_pic} they are compared
to the exact values, to the second order result and to the values coming from 
the direct application of bosonization (cf.~appendix C).
Clearly, the CUT results agree very well with exact data for not too
large values of $V < 0.6$. From the comparison to the 
exact second order term $V^2$ one sees that the CUT data describes
the full exact result better than the second order
term alone. We infer that the CUT data reproduces the third order term also. 
This does not come as a surprise since we showed above that the
ground state energy is exact including the $V^3$ term. Thus any quantity
derived from it will also be exact in the same order. So the momentum
distribution $n(k)$ and hence the exponent $\alpha$ have to be exact 
 up to and including $V^3$.

The comparison to the bosonization result in Fig.~\ref{alpha_pic}
is also instructive. By construction the bosonization result
captures only the physics in an infinitesimal vicinity of the
Fermi points. There no umklapp scattering is possible as
can be seen from Eq.~(\ref{mg}) or from the analyses in Refs.\
\cite{shank91,shank01} (cf.\ discussion after Eq.~(\ref{r1})).
For this reason, the bosonization fails to detect any precursor
of the incipient phase transition to the CDW occurring at $V=1$.
The dependence of the $\alpha$ obtained from bosonization
 is hence much smoother
than the exact result. The CUT result includes scattering at all
momenta. So precursors of the phase transition can be captured.
Unfortunately, the breakdown of the approximation as used here
makes further statements on the description of the phase transition
impossible.

\section{Discussion}

\subsection{Conclusions}

We presented general arguments in favour for 
a renormalization treatment of interacting fermionic systems
 by means of a continuous unitary transformation (CUT). 
The precise choice of the unitary transformation, i.e.\
the choice of its infinitesimal generator, was motivated and the
general properties of the transformation were elucidated.
The method was illustrated for a model of one-dimensional, 
repulsively interacting fermions without spin at half-filling.
The technical considerations for the explicit calculation
were given. An important point was a consistent, non-redundant
 notation (\ref{r1}) which made the crucial cancellation
between the Cooper pair and the zero sound channel manifest.
 Results were obtained for the correlation part
of the ground state energy, for the 1-particle dispersion, for the
2-particle vertex function and for the static momentum distribution.
The findings were compared to exact results as far as possible.
The agreement was very good.

The CUT employed uses standard quasi-particles as elementary
excitations. So it represents an explicit construction of
Landau's mapping of the non-interacting quasi-particles to the
elementary excitations of the interacting system \cite{heidb02a}. 
Note that the existence of such a smooth connection is not
surprising since already the bosonization identity for fermionic
field operators \cite{halda81b,voit95} represents such a smooth
link. In the continuum limit, a spinless model \emph{with} or
\emph{without} interaction can be mapped to a single mode boson
model with linear dispersion. Hence, it is possible to link
the interacting model via the bosonic model to the non-interacting
one (see also the remark on Kehrein's results \cite{kehre99,kehre01} below).
We take the success of our  approach as corroborating
(numeric) evidence that 
Landau's mapping exists in one dimension. In particular,
the numerical findings of the momentum distribution $n(k)$ indicate that 
important features of Luttinger liquids could be retrieved. Signatures
of a Luttinger-type power law behaviour at the Fermi points were found,
%%% NEU
though hampered by the accessible restricted system sizes. 
%%% NEU

We are convinced 
 that  CUTs represent a powerful renormalization scheme for low-dimensional
 systems. Since no states are eliminated the effective hamiltonian
obtained at the end of the CUT allows to compute spatial (shown here) 
and  temporal correlations (for an example, see Ref.~\cite{knett01b})
at small \emph{and} at large wave vectors or excitation energies, 
respectively. 
So, in principle, no information is lost in the course of the renormalization,
in contrast to, for instance, Wilson's renormalization \cite{wilso75}.
Of course, approximations which are necessary in practical calculation
will introduce some uncertainties.
%%% NEU 
Recent developments in renormalization approaches
 by integrating out degrees of freedom
allow also to compute high energy features if the observables
are equally subject to the flow, see for instance the appendix 
in Ref.~\cite{salmh01}. A major advantage of the CUT approach is
that no frequency dependence needs to be kept. This represents 
an important facilitation for the actual numerical  realization of the
renormalization.

\subsection{Connections to other Work}

\paragraph{Life-Time of Excitations} If  a complete or partial 
diagonalization is obtained by a continuous unitary transformation
the  eigen values are by construction real. So the excitations
are well-defined in energy and they do not display a finite life-time.
This statement appears almost trivial if one bears  in mind
only the mathematical linear algebra. From the physics points of view, however,
one might be surprised since one is  used to that
 excitations, for instance quasi-particles, have
a finite life-time in many-body systems. This comes about because
no true eigen state are considered when an excitation
of finite life-time is studied. 

For instance, adding a fermion to the ground
state of an interacting fermion system by application of a simple
creation operator generically does not yield an eigen state but a
sophisticated superposition of true eigen states. This looks as if
the ``true'' fermion added decayed because its spectral function
displays a peak of some finite width. Technically, one still uses
the normal single-particle basis but the self-energy $\Sigma(k,\omega)$
 acquires a finite imaginary part which represents the fact that the single
fermion is coupled to states with one or more particle-hole pairs. 
One may view $\varepsilon_k +\Sigma(k,\omega)$ as an imaginary
 eigen value of the single particle state. But the single particle
state is not an eigen state of the underlying hamiltonian.

The description using a unitary transformation designed to 
diagonalize the original hamiltonian is different from the
standard approach in physics sketched above. One tries to
find real eigen vectors and eigen values. Broad spectral
functions do not occur because of imaginary eigen values but
because of the superposition of eigen vectors. Kehrein and
Mielke coined the expression that ``the observables decay'' and
described the phenomenon in the context of dissipation \cite{kehre97}.

So it is not surprising that, after the CUT is applied to one-dimensional
fermions, we find  quasi-particles with \emph{infinite} life-time. These are 
quasi-particles \emph{after} the transformation. The fermion \emph{before}
the transformation will decay into states with additional 
excited particle-hole states. So there is no contradiction.
In this context it is interesting to note that there is a formulation
of diagrammatic perturbation theory which uses also infinite life-time
excitations \cite{balia61a,balia61b}. In this formulation the fermionic
one-particle states
 acquire a renormalized eigen energy due to the interaction. The change of
the eigen energy $\varepsilon \to \varepsilon'$
leads to a change in the occupation number 
$(1-\exp(-\beta(\varepsilon-\mu)))^{-1}\to 
(1-\exp(-\beta(\varepsilon'-\mu)))^{-1})$ \cite{balia61a}. So this
approach corroborates  that a description of the physics 
in terms of non-decaying quasi-particles is possible. The question
how dynamic correlations can be described is not discussed in
Refs.~\cite{balia61a,balia61b}.

\paragraph{CUTs as Numerical Approach}
While the present work was being 
finished White \cite{white02} proposed a numerical 
 scheme which is very similar in spirit to what we did here. Besides 
discrete transformations which work less efficiently, a continuous
unitary transformation is used with an infinitesimal generator 
with matrix elements
\begin{equation}
  \eta_{ij}(\ell)=\frac{1}{E_{i}(\ell)-E_{j}(\ell)} H_{ij}(\ell)\ ,
\label{white}
\end{equation}
where a 1-particle basis is used, i.e.\ a basis in which
the 1-particle part of the hamiltonian is diagonal. The 
1-particle eigen energies are given by $E_i$. The approach is applied
to a small molecule, namely H$_2$O, and the ground state energy is
computed very reliably by rotating the original ground state to
a Fermi sea, i.e.\ the quasi-particle vacuum. 
That means that states which partially occupied $n\geq 0.5$
are mapped to filled states and states which are partially empty
are mapped to empty states. This is what we did in our present work
as well. So Ref.~\cite{white02} provides an
independent investigation of the power of continuous unitary 
transformations for a 
%%% NEU kein completely
 different fermionic system.
 In addition, it is investigated in Ref.~\cite{white02}
to eliminate a number of states which lie far off the Fermi level without
 diagonalizing the problem completely by the continuous transformation.
The remaining effective problem, which is simplified considerably due
to the reduction of the Hilbert space, is then solved by standard
diagonalization algorithms, e.g.\ DMRG. Also this approach proved to
be very powerful.  

\paragraph{Choice of CUT}
In Ref.~\cite{wegne94}, Wegner investigated a one-dimensional $n$-orbital model
in the continuum limit by a continuous unitary transformation. 
The main difference in the unitary transformation is the use of
a generator different from the one in
Eq.~(\ref{mku_cuts}), namely 
\begin{equation}
\label{wchoice}
\eta = [H_{\rm D},H]
\end{equation}
 where $H_{\rm D}$
is the part of the Hamilton one wants to keep. The approach succeeded
when $H_{\rm D}$ comprised all terms that do not change the number
of quasi-particles. Hence this renormalizing scheme is very similar to
the one used in the present work.
It would be an interesting issue to
compare both approaches quantitatively in a simple model. There are
arguments in favour for both of the two approaches.

First, the choice (\ref{mku_cuts}) has the advantage that the kind of 
terms that are generated is restricted: the block band structure is preserved.
Second, off-diagonal terms changing the number of quasi-particles
are eliminated even if they are not accompanied by a change of the
1-particle energies, i.e.\ certain degeneracies are lifted.
Third, a suppression of off-diagonal parts starts already linearly
in the differences of the diagonal parts.
A weakness of the choice (\ref{mku_cuts}) is given when there are
scattering processes which increment the number of quasi-particles
but decrease the 1-particle energies. In this case, the corresponding
amplitudes are first enhanced before the decrease to the end of the
transformation. Due to the necessary truncations it may be difficult
to control the quality of the approximation during the stage of 
enhancement.

On the other side, the choice (\ref{wchoice}) is firstly very robust
since off-diagonal terms are always suppressed due to the fact
that the energy difference occurs squared \cite{wegne94}. Second, one
does not need to know explicitly the eigen basis of $H_{\rm D}$. The
squares are generated automatically by the double commutator when
Eq.~(\ref{wchoice}) is combined with Eq.~(\ref{general_fe}). Yet these
two commutators must also to be computed which might be tedious.
Another weakness arises when large-scale degeneracies spoil the method
by stopping the renormalization prematurely. For instance, the vanishing of
$\eta$ implies the stop of the flow but guarantees only that there is
a common basis set of $H_{\rm D}$ and $H$, not that $H$ is diagonal.
So the conservation of the number of quasi-particles is not ensured
by the choice (\ref{wchoice}).

Summarizing the comparison of the choices (\ref{mku_cuts},\ref{wchoice})
we reckon that (\ref{mku_cuts}) works better if the number of
excitations correlates well with the energy. If this is not the case, 
the robustness of (\ref{wchoice}) may be preferable. Note that there
are still completely different generators conceivable \cite{grote02}. 
For example,  one can use in a basis of 1-particle states
\begin{eqnarray}
  \eta_{ij}(\ell)&=& 0 \quad \mbox{if}\ q_i=q_j \nonumber\\
 \eta_{ij}(\ell)&=&\mbox{sign}(E_{i}-E_{j})H_{ij}  \quad \mbox{otherwise}
\label{neu}
\end{eqnarray}
where in case of degeneracy $E_i=E_j$ the energy of the state with more
excitations is assumed to be infinitesimally 
higher. Such an approach captures
advantages of (\ref{mku_cuts}) while avoiding its disadvantage.
Further investigations
of these issues are certainly called for.  

\paragraph{Fermionic Excitations}
The fact that we treat a system of interacting one-dimensional fermions
in its Luttinger liquid phase without using explicitly collective bosonic
modes might be surprising. Yet it is not uncommon that an interacting
system is suitably described (after certain transformations) 
by free or nearly free fermions. As an example we quote the work by
Kehrein who succeeded to map a sine-Gordon model by a sequence of
continuous transformations onto a model of free fermions 
\cite{kehre99,kehre01}. Indeed, Kehrein's mapping accomplishes
this aim for a broader range of parameters than previous renormalization
treatments. 
%%% NEU
The neglect of interactions
between the excitations viewed as elementary is a fairly severe 
approximation for $\beta^2 < 4\pi$ where the breathers, i.e.\
bound states, are known to occur \cite{bergk79}. Their description
requires the inclusion of the interaction between elementary
excitations.

\paragraph{Landau's Fermi Liquid}
Finally, we wish to comment on the use of a Landau's Fermi
liquid description in terms of quasi-particles for one-dimensional
systems \cite{heidb02a}. The possibility and the power of such a description 
has been noted previously by Carmelo et al.\ 
%\cite{carme91a,carme92a,carme92b,carme92c} 
\cite{carme}
in the framework of the Bethe
ansatz solution of the one-dimensional Hubbard model. Carmelo and coworkers
use the spinons and holons  as they arise  in the
Bethe ansatz solution as pseudo-particles. Then an approximate treatment
for the spatial and temporal \cite{carme00} correlations is built by 
describing the  excitations as small deviations from the ground state
distributions of these pseudo-particles. In this sense, the concept
of Landau's Fermi liquid is generalized to one dimension. The author
emphasize, however, that the pseudo-particles cannot be smoothly
linked to the \emph{quasi}-particles of the non-interacting solution.
In this point, a clear difference to our finding here and in 
Ref.~\cite{heidb02a} occurs. We argue on the basis of our
numerical results that a smooth mapping between
the interacting and the non-interacting excitations exists even
in one dimension. 
%%% NEU
The existence of such a mapping as long as the
system remains massless can already be deduced from bosonization.
Since the interacting and the non-interacting model can be mapped
to a model of free linear dispersion bosons (discarding Umklapp
scattering) they can also be mapped to each other.
By means of the CUT we constructed such a mapping explicitly.

The generalized Landau liquid in 
Refs.~\cite{carme}
%\cite{carme91a,carme92a,carme92b,carme92c} 
relies on the Bethe
ansatz solution of the one-dimensional Hubbard model. Hence it may
be that the integrability is a prerequisite for the generalized Landau liquid.
In our calculation in contrast, 
the integrability of the model studied does not play
a r\^ole other than providing a rigorous benchmark.
%%% NEU
But so far, we have not considered the spinful case.
Its investigation by CUTs is certainly called for.

\subsection{Outlook}
A comprehensive summary is given in Sect.~4.1. 
Here we point out in which directions
further work is required. In view of the numerical nature of the
present work, an  analytical treatment would be helpful. 
There is still a certain gap between the analytical result of a
logarithmic divergence for the momentum distribution 
obtained in Ref.~\cite{wegne94} and the numerical results we found.
For an analytical treatment the models to be considered have to be
simplified further. Spin, however, should be included in order
to enlarge decisively  the class of systems which can be described.

Another very interesting issue is the computation of dynamical
quantities like the local spectral function $A(\omega)$ or the
momentum resolved spectral function $A(k,\omega)$. In the framework of
 standard renormalization such investigations are presently
 carried out \cite{busch02}. The dynamical quantities
are of interest to see theoretically to which extent and to which accuracy 
they can be computed at all energies and momenta.
For the explanation
of experimental data the spectral functions are of utmost importance.
It is this objective which requires in particular to go beyond 
the asymptotic regime of very small energies and momenta
\cite{claes02}. For gapful spin systems dynamical quantities have already
been computed successfully \cite{knett01b}. The results agree very well with
experiments and render deeper insight in the underlying physics
\cite{windt01,schmi01}.

In order to go beyond one-dimensional systems, modified
generators have to be investigated, see e.g.~\cite{grote02}. The
pros and cons of the choices used presently were briefly discussed
in the preceding section. The issue of the optimum generator
represents a longer-lasting question since the answer depends
certainly on the model to be studied.

\paragraph{Acknowledgements}
It is a pleasure for us to acknowledge interesting and helpful discussions
with N.~Grewe, S.~Kehrein, 
P.~Horsch, E.~M\"uller-Hartmann, M.~Salmhofer, S.R.~White and P.~W\"olfle.
This work has been supported by the Deutsche Forschungsgemeinschaft
through SP 1073 and through SFB 608.

% BibTeX users please use
%\bibliographystyle{prsty}
%\bibliography{../bibinput/liter10}

\begin{appendix}

\section{Normal-Ordering}
Normal-ordering \cite{wick50} denoted by colons $:A:$
is a standard procedure which is explained
at length in the text books \cite{negel88}. So we recall here
only the gist of it which is technically relevant for our calculation.
For further details we refer the reader to the concise script by
Wegner \cite{wegne00}.

Considering a product $X_{n}$ of $n$ fermionic operators it is a priori
not clear on how many particles this operator really acts. It looks as
if it acted on $n$ fermions. But a part of this action may be
redundant in the sense that it can be expressed also by an operator
$Y_m$ with $m<n$. This is indeed the generic situation.
 A point of reference is needed in order to be able to define how
many particles are involved in a certain process. Given the
ground state of a 1-particle hamiltonian\footnote{The formalism
works identically for finite temperatures  with respect
to the statistical operator of a 1-particle hamiltonian.}
the normal-ordering ensures that the normal-ordered $n$ operator
does not contain parts which can be viewed as action of 
an operator with less fermionic factors.
This is the physical content of 
\begin{equation}
\langle :P_n: :Q_m:\rangle =0\quad \mbox{for} \quad n\neq m \ 
\end{equation}
as derived for normal-ordered terms \cite{wegne00}.

Technically, a usual product of $m$ fermionic operators
$a_k$ is expressed in terms of normal-ordered terms as
\begin{eqnarray}
\nonumber
&&a_{k1}a_{k2} \ldots  a_{km}  = \\
\label{hin}
&&\quad 
:\exp\left(\sum_{k,l} G_{kl} 
\textstyle\frac{\partial^2}{\partial a_l^{\rm right}
\partial a_l^{\rm left}} \right) a_{k1}a_{k2} \ldots  a_{km} :\ , \quad
\end{eqnarray}
where $G_{kl}$ is the contraction $\langle a_k a_l\rangle$.
The superscripts `left' and `right'
 indicate that in the double derivatives only
pairs are taken where $a^{\rm left}_k$ is a factor to the left of
$a^{\rm right}_l$. To obtain the correct signs the `left'
derivation must be taken before the `right' excitation.
Eq.~(\ref{hin}) stands for the known procedure that a product is normal
ordered by writing down the sum of terms with all possible numbers
and sorts of contractions. The inverse relation is given simply by
\begin{eqnarray}\nonumber
&& :a_{k1}a_{k2} \ldots  a_{km}: = \\
\label{her}
&&\quad
\exp\left(-\sum_{k,l} G_{kl} 
\textstyle\frac{\partial^2}{\partial a_l^{\rm right}
\partial a_l^{\rm left}} \right) a_{k1}a_{k2} \ldots  a_{km} \ . \quad
\end{eqnarray}
Combining Eqs.~(\ref{hin}) and (\ref{her}) leads to the useful expression
for products of normal-ordered terms $:A(a):$ and $:B(a):$
\begin{equation}
\label{product}
:A(a): :B(a): = 
:\exp\left(\sum_{k,l} G_{kl} 
\textstyle\frac{\partial^2}{\partial b_l
\partial a_l} \right) A(a) B(b) :\Big|_{b=a}
\end{equation}
where the superscripts `left' and `right' are no longer needed
due to the sequence of factors in the product.
In practice, Eq.~(\ref{product}) means that for the normal-ordering 
of a product of already normal-ordered factors not all contractions
need to be considered. Only those contractions matter where the two 
fermionic operator
do not come from the same factor. This is easy to understand since
the contractions between fermionic operators from the same factor
are already accounted for by the normal-ordering of each factor
separately.

In order to determine the differential equations resulting from
Eq.~(\ref{general_fe}) commutators of normal-ordered terms must
be computed. To do this \emph{without}
 passing by non-normal-ordered expressions an extension of
Eq.~(\ref{product}) to commutators is particularly useful.
We derived and checked the identity
\begin{eqnarray}\nonumber
&&[:A(a):, :B(a):] = \\ \nonumber
&& \qquad:\exp\left(\sum_{k,l} G_{kl} \left(
\textstyle\frac{\partial^2}{\partial a_l^{\rm right} 
\partial b_k^{\rm left}} + 
\textstyle\frac{\partial^2}{\partial b_l^{\rm right} 
\partial a_k^{\rm left}} \right) \right) \\
&&\qquad\qquad
[A(a),B(b)] :\Big|_{b=a}\ ,\quad 
\label{commutator}
\end{eqnarray}
where the commutator is computed using the anticommutators
$\{ a_k, b_l \} := \{ a_k, a_l\}$.
Eq.~(\ref{commutator}) means that one can first compute the commutator
as usual, but remembering whether the fermionic operator comes from 
$A$ or from $B$. Then normal-ordering is achieved by writing down
the terms with all possible contractions between pairs of fermionic operators
where one comes from $A$ and the other from $B$.
In this way, the computation of the actual general flow equation
(\ref{general_fe}) becomes a task which is not too demanding.

\section{The Self-Similar CUTs}
\label{derivation}
In the self-similar  approxmation we compute
\begin{equation}
\left[T^{(0)}_{0}(\ell)+T^{(1)}_{0}(\ell)+\sum^{+2}_{-2}T^{(2)}_{2i}(\ell),
\sum^{+2}_{-2}\mbox{sign}(i)T^{(2)}_{2i}(\ell)\right]
\end{equation}
neglecting the arising $T^{(3)}_i$ terms. 
The change in the number of quasi-particles for the expression\\
 $\Gamma_{kqp}:
c^{\dagger}_{k+q}c^{\dagger}_{k-q}c^{\phantom\dagger}_{k-p}
c^{\phantom\dagger}_{k+p}:$ is given by $S_{kqp}$ as defined by
\begin{eqnarray}
S_{kqp}&=&\mbox{sign}(n_{k+p}+n_{k-p}-n_{k-q}-n_{k+q})\ ,
\end{eqnarray}
where we use $n_k$ for the momentum distribution of the
 unperturbed Fermi sea since this is the quasi-particle vacuum to
which we are mapping the ground state. (The actual momentum
distribution is denoted $n(k)$).
With these definitions we calculate the 
commutator using normal-ordering as explained in appendix A.
Then we compare the coefficients of the various terms
(0-particle, 1-particle and 2-particle terms) and determine in this
way the set of differential equations.

\subsection{Ground state energy per site}
The differential equation of the ground state energy per site depends 
only on the commutator $[T^{(2)}_{\pm 4},T^{(2)}_{\mp 4}]$ where all fermionic 
operators are contracted. There are four different ways to combine the 
operators for the complete contractions. 
But all of them lead to the same expression if one uses the
 symmetries Eq.(\ref{s1})-(\ref{r1}). Finally one obtains
\begin{eqnarray}
&&\frac{d}{dl}\frac{E}{N}=
\frac{8}{N^3}\!\!\!\! \sum_{{k\in[0,\pi)}\atop{qp\in[-\pi,\pi)}}\!\!\!\!
(1-2\,n_{k+p})\,n_{k+q}\,n_{k-q}\,S_{k,q,p}\,{\Gamma_{k,q,p}^2} .
\qquad
\label{mku_E}
\end{eqnarray}

\subsection{Dispersion}
For the 1-particle term, the dispersion, one has to take 
all combinations of three contractions into account that occur 
 in $[T^{(2})_i,T^{(2)}_{-i}]$ with $i\in\{0,\pm2,\pm4\}$. 
One obtains $16$ different parts that turn out to be identical.
In order to avoid double-counting one of the free momenta must
be restricted to $[0,\pi)$ or the sum must be divided by two. 
So one obtains finally
\begin{eqnarray}
&&\frac{d}{dl}\varepsilon_{k}=\nonumber\\
&&\quad\frac{8}{N^2}\!\!\!\!\sum_{qp\in[-\pi,\pi)}\!\!\!\!
((1-2n_{k+q-p})n_{k-2q}+n_{n+p-q}n_{n-p-q})\times\nonumber\\
&&\qquad S_{k-qqp}\Gamma^2_{k-qqp} \ .
\label{mku_e}
\end{eqnarray}

\subsection{Vertex Function}
For the vertex function $\Gamma_{kqp}$ there are two commutators to calculate: 
$[T^{(2)}_i,T^{(2)}_j]$ ($i,j\in\{0,\pm2,\pm4\}$) with all possible 
combinations of two contractions  and  $[T^{(1)}_0,T^{(2)}_j]$ with all 
possibilities of a single contraction. 
To keep the notation short, we define
\begin{equation}
\Phi_{kqp|KQP}:=(S_{kqp}-S_{KQP})\Gamma_{kqp}\Gamma_{KQP}\ .
\end{equation}
The differential equation for the flow of $\Gamma_{kqp}$ 
then reads
\begin{eqnarray}
&&\frac{d}{dl}\Gamma_{kqp}=
(\varepsilon_{k+p}+\varepsilon_{k-p}-\varepsilon_{k-q}-
\varepsilon_{k+q})S_{kqp}\Gamma_{kqp}+ \nonumber\\
&&\quad\frac{1}{N}\sum_{Q\in[-\pi,\pi)}4n_Q \Big\{\nonumber\\
&&\quad\Phi_{\frac{k+q+Q}{2}\frac{k+q-Q}{2}\frac{k-q-Q}{2}-p|\frac{k+p+Q}{2}
\frac{k-p-Q}{2}-q\frac{k+p-Q}{2}}+\nonumber\\
&&\quad\Phi_{\frac{k-q+Q}{2}\frac{k-q-Q}{2}\frac{k+q-Q}{2}+p|\frac{k-p+Q}{2}
\frac{k+p-Q}{2}+q\frac{k-p-Q}{2}}-\nonumber\\
&&\quad\Phi_{\frac{k+q+Q}{2}\frac{k+q-Q}{2}\frac{k-q-Q}{2}+p|\frac{k-p+Q}{2}
\frac{k+p-Q}{2}-q\frac{k-p-Q}{2}}-\nonumber\\
&&\quad\Phi_{\frac{k-q+Q}{2}\frac{k-q-Q}{2}\frac{k+q-Q}{2}-p|\frac{k+p+Q}{2}
\frac{k-p-Q}{2}+q\frac{k+p-Q}{2}}\nonumber\\
&&\quad\Big\}+2(1-2n_Q)\Phi_{kq(Q-k)|k(Q-k)p}\ .\label{mku_g}
\end{eqnarray}
Note that the appearance of {\em four} $\Phi$-terms is due to the fact that
we denote the scattering processes in the hamiltonian (\ref{general_ham})
in a notation symmetric way (\ref{r1}).  A naive
comparison of coefficients would lead only to one of the four terms.
The other three come into play if one requires that 
$\frac{d}{dl}\Gamma_{kqp}$ fulfills (\ref{r1}).
As explained in the main text, the notation obeying (\ref{r1})
ensures that a maximum number of cancellations are dealt
with explicitly. This is advantageous on the numerical as well as on the
conceptual level. 

\section{Bosonization of the Model}
\label{bosonization}
We employ a constructive bosonization by linearizing the dispersion on both
 branches ($r=+1\leftrightarrow$ right branch, $r=-1\leftrightarrow$ 
left branch)
\begin{equation}
\varepsilon_{k,r}=r\left(1+\textstyle\frac{2V}{\pi}\right)(k- r k_{\rm F})
\ .
\end{equation}
Furthermore, the bare interaction vertex (\ref{mg}) 
is evaluated with all momenta
being taken at the Fermi points $\pm k_{\rm F}$.
The corresponding scattering strengths take the value $\pm V$. 
Following Ref.~\cite{voit95} one  can then determine the 
renormalized Fermi velocity as
\begin{equation}
v^*_{\rm F}(V)=\left(1+\textstyle\frac{2V}{\pi}\right)
\sqrt{1-\left(\textstyle\frac{2V}{\pi+2V}\right)^2}
\end{equation}
and the exponent $\alpha$ occurring in the momentum distribution from
Eq.~(\ref{alpha-eta}) and
\begin{eqnarray}
%\alpha(V)&=&\frac{1}{\eta_0(V)}+\frac{\eta_0(V)}{4}-1\nonumber\\
\eta_0(V)&=&\frac{1}{2}\sqrt{\frac{2\pi}{2\pi+8V}}\ .
\end{eqnarray}
\end{appendix}

\end{document}